\documentclass[fleqn,usenatbib]{mnras}
\usepackage{newtxtext,newtxmath}
\usepackage[T1]{fontenc}
\DeclareRobustCommand{\VAN}[3]{#2}
\let\VANthebibliography\thebibliography
\def\thebibliography{\DeclareRobustCommand{\VAN}[3]{##3}\VANthebibliography}
\usepackage{graphicx}	
\usepackage{amsmath}

\usepackage{hyperref}
\usepackage{orcidlink}

\hypersetup{
    colorlinks=true,
    linkcolor=blue,
    filecolor=blue,
    urlcolor=blue,
    citecolor=blue,
    pdfborder={0 0 0}
}

\title[Stellar Splashback in Clusters]{The Three Hundred Project: deducing the stellar splashback structure of galaxy clusters from their orbiting profiles}

\author[K. Walker et al.]{K. Walker$^{1,2}$\textsuperscript{\href{https://orcid.org/0000-0003-2128-1289}{\orcidlink{0000-0003-2128-1289}}},
\thanks{E-mail: kris.walker@icrar.org}
A. Ludlow$^1$\textsuperscript{\href{https://orcid.org/0000-0001-6119-4871}{\orcidlink{0000-0001-6119-4871}}},
C. Power$^{1,2}$\textsuperscript{\href{https://orcid.org/0000-0002-4003-0904}{\orcidlink{0000-0002-4003-0904}}},
A. Knebe$^{1,3}$\textsuperscript{\href{https://orcid.org/0000-0003-4066-8307}{\orcidlink{0000-0003-4066-8307}}}, \&
W. Cui$^3$\textsuperscript{\href{https://orcid.org/0000-0002-2113-4863}{\orcidlink{0000-0002-2113-4863}}}
\\
$^1$ International Centre for Radio Astronomy Research, The University of Western Australia, 35 Stirling Highway, Crawley, Western Australia, 6009, Australia\\
$^2$ ARC Centre of Excellence for All Sky Astrophysics in 3 Dimensions (ASTRO 3D)\\
$^3$Departamento de Física Teórica, Módulo 15, Facultad de Ciencias, Universidad Autónoma de Madrid, 28049 Madrid, Spain}

\date{Accepted 2026 February 16. Received 2026 January 28; in original form 2025 August 10}

\pubyear{\the\year{}}

\begin{document}
\label{firstpage}
\pagerange{\pageref{firstpage}--\pageref{lastpage}}
\maketitle

\begin{abstract}
We examine the splashback structure of galaxy clusters using hydrodynamical simulations from the GIZMO run of The Three Hundred Project, focusing on the relationship between the stellar and dark matter components. We dynamically decompose clusters into orbiting and infalling material and fit their density profiles. We find that the truncation radius $r_{\mathrm{t}}$, associated with the splashback feature, coincides for stars and dark matter, but the stellar profile exhibits a systematically steeper decline. Both components follow a consistent $r_{\mathrm{t}}{-}\Gamma$ relation, where $\Gamma$ is the mass accretion rate, which suggests that stellar profiles can be used to infer recent cluster mass growth. We also find that the normalisation of the density profile of infalling material correlates with $\Gamma$, and that stellar and dark matter scale radii coincide when measured non-parametrically. By fitting stellar profiles in projection, we show that $r_{\mathrm{t}}$ can, in principle, be recovered observationally, with a typical scatter of $\sim 0.3\,R_{200\mathrm{m}}$. 
Our results demonstrate that the splashback feature in the stellar component provides a viable proxy for the cluster's physical boundary and recent growth by mass accretion, offering a complementary observable tracer to satellite galaxies and weak lensing.
\end{abstract}




\section{Introduction}
\label{sec:intro}

The precise definition of a galaxy cluster’s outer boundary is fundamental to studies of cluster formation, evolution, and cosmology \citep[e.g.][]{Kravtsov.Borgani.2012,Morandi.Sun.2016,Haggar.etal.2024}. Traditional definitions---deduced from friends-of-friends (FoF) linking lengths  or spherical overdensity radii---play a key role in determining cluster masses and membership \citep[e.g.][]{Eke.etal.2004,Lukic.etal.2009}, yet these are choices driven primarily by what is practicable with data rather than what is most physically meaningful \citep[e.g.][]{White.2001,Old.etal.2014,Old.etal.2015}. Establishing a boundary that is physically meaningful and practicably measurable via observations remains a major challenge.

A promising physically motivated boundary is provided by the splashback radius, which marks the apocentre of recently accreted material orbiting within the cluster potential (e.g. \citealt{Diemer2014, Adhikari2014, More2015, Diemer2022}). This radius delineates the transition between orbiting and infalling matter, and manifests as a steepening in the halo density profile (which, we note, has historically---and erroneously---been conflated with the splashback radius itself). The radius of this steepest slope is of particular interest because there is now considerable evidence that this steepening has been measured in observations of real galaxy clusters (e.g. \citealt{More2016, Baxter.etal.2017, Nishizawa2018,Chang2018, Contigiani.etal.2019, Xu2024}). This is exciting because there is good reason to expect that splashback radius offers a direct probe of halo assembly history and mass accretion rates, thereby connecting cluster outskirts to cosmological growth \citep{More2015,Deason2021,Dacunha2025}.

Observational detections of splashback features have been reported using satellite galaxy counts \citep[e.g.][]{More2016, Baxter.etal.2017, Nishizawa2018,Adhikari.etal.2021} and weak gravitational lensing \citep[e.g.][]{Chang2018, Contigiani.etal.2019, Xu2024,Giocoli2024}. Initial observations suggested that the splashback radius is smaller than that predicted by $\Lambda{\rm CDM}$ simulations by about $20\%$ \citep{More2016,Baxter.etal.2017,Chang2018}. It has since been argued that this is the result of biases arising from projection effects in optically-selected clusters \citep{Busch2017,Sunayama2020}, particularly when using apertures significantly smaller than the splashback radius \citep{Murata2020}. These biases have been mitigated by better optical identification \citep{Murata2020}, or avoided altogether by selecting clusters by their Sunyaev-Zel'dovich (SZ) signal \citep{Shin.etal.2019, Zurcher.More.2019,Adhikari.etal.2021} or X-ray luminosity \citep{Rana2023,Joshi.etal.2026}.

Methods for detecting the steepening feature have their own limitations. In the case of satellite galaxy counts, accurate measurement of splashback features depends on the properties of satellite tracers in a given cluster \citep[e.g.][]{ONeil.etal.2022}, and accounting for the complexity of their kinematics \citep[e.g.][]{Aung.etal.2021} and structure of the local cosmic web \citep[e.g.][]{Lebeau.etal.2024,Sun.etal.2025}. Satellite galaxies are also biased tracers of the cluster potential, with the most massive tending to accumulate towards the central regions due to dynamical friction \citep{Ludlow2009,Adhikari2016}. Weak lensing directly probes the total projected mass profile and is free of such biases, but its signal-to-noise ratio is typically low because individual systems contain only a small number of lensed background sources.

A promising alternative probe is the diffuse intra-cluster light (ICL) that is formed from the remnants of disrupted galaxies. This material is a potentially powerful tracer of cluster outskirts \citep{Montes2019,Contini2021,ContrerasSantos2024}, and a potential detection of the steepening feature in the ICL of an individual cluster has already been reported by \citet{Gonzalez2021}. Theoretical work by \citet{Deason2021} has shown a direct correspondence between the steepening feature in the total stellar and dark matter density, indicating a dynamical link between these components. However, there remain important questions about the detailed nature of the stellar splashback structure, its relation to the underlying dark matter, and its observational accessibility, which we seek to answer in this paper. 

A deeper understanding of the stellar splashback feature is crucial for a number of reasons. The distribution of stars in a cluster reflect physical processes that affect satellite galaxies and their dark matter halos, such as tidal stripping \citep{Sifon2018,Kumar2026,Xie2015} and dynamical friction \citep{Ludlow2009,Adhikari2016}, and so their distribution encodes information about the interplay between baryons and dark matter in the cluster outskirts \citep[e.g.][]{Brown.etal.2024}. When combined with splashback and shock structures in the stars and gas \citep[e.g.][]{Deason2021,Towler.etal.2024,Zhang.etal.2025}, this information can be used to infer dynamical history. Similarly, projected stellar density profiles provide a complementary means of constraining cluster growth histories \citep[e.g.][]{Caminha.etal.2017,Montes.Trujillo.2018,Kluge.etal.2025,Joshi.etal.2026}. This is especially valuable when supplemented by weak lensing; combining different datasets, affected by projection in different ways, allows for a more robust inference of splashback features \citep[e.g.][]{Adhikari.etal.2021}.

In this paper, we build upon recent work by \citet{Diemer2022,Diemer2023,Diemer2025} and investigate the splashback region surrounding galaxy clusters by distinguishing material that is orbiting within the halo from material that is infalling for the first time. To do this, we use the statistical sample of massive galaxy clusters from {\small The Three Hundred} collaboration's hydrodynamical galaxy formation simulation suite \citep[cf.][]{Cui.etal.2018,Cui.etal.2022}. This is a mass complete sample of clusters drawn from a 1 $h^{-1}$ Gpc box, which have a diversity of assembly histories and larger-scale environments. These simulations self-consistently model the physics of galaxy formation, including star formation, black hole growth, supernovae, and AGN feedback. By investigating the stellar splashback structure and its relation to dark matter in hydrodynamical simulations, our aim is to establish a physically grounded framework for interpreting stellar density profiles as tracers of cluster assembly. This approach promises to enhance the utility of stellar observables---such as the ICL---in constraining cluster boundaries and mass accretion histories, thereby broadening the toolkit available for cluster cosmology.

The remainder of this paper is organised as follows. \S\ref{sec:data} summarises the simulation data used in our analysis, and \S\ref{sec:methods} describes our approach to decomposing and characterising the galaxy cluster structure into orbiting and infalling components. In \S\ref{sec:results} we present our key results, and we discuss these results in the context of previous studies in \S\ref{sec:discussion}. Finally, we summarise our key results and identify future directions in \S\ref{sec:conclusions}


\section{The data}
\label{sec:data}
For this work we use data from the \textsc{gizmo-simba} run of the \textsc{Three Hundred Project} \citep{Cui.etal.2018}, a suite of hydrodynamical re-simulations of the most massive 324 galaxy cluster halos in the dark matter-only MultiDark Planck 2 (MDPL2) simulation \citep[cf.][]{Klypin.etal.2016}, a $1 h^{-1}$ Gpc box on a side. These clusters have virial masses in the range $6.4 \times 10^{14} h^{-1}M_{\odot} \lesssim M_{200{\rm c}} \lesssim 2.6 \times 10^{15} h^{-1}M_{\odot}$, where $M_{200{\rm c}}$ is the mass corresponding to an overdensity criterion of 200 times the critical density at that epoch. Each re-simulated region extends $15\,h^{-1}\,{\rm Mpc}$ in comoving radius from the $z=0$ cluster centre, corresponding to several virial radii.

The re-simulations were run using the meshless finite mass (MFM) implementation of the \textsc{gizmo} hydrodynamics code \citep[cf.][]{Hopkins.2015}. This method overcomes some of the limitations of classic smoothed-particle hydrodynamics and moving-mesh approaches and captures fluid mixing and instabilities. The gravity solver used in \textsc{gizmo} is derived from \textsc{gadget-3} \citep[][]{Springel.2005}. Galaxy formation physics is implemented using a modified version of the \textsc{simba} model \citep[][]{Dave.etal.2019} that has been calibrated for cluster scales \citep[cf.][]{Cui.etal.2022}. It models radiative cooling, star formation and feedback, black hole formation and growth, and multiple modes of black hole feedback. Star particles are spawned from gas elements stochastically, with each MFM fluid element spawning a single star particle of the same mass. The \textsc{gizmo-simba} model was run on re-simulation regions from both a $3840^3$- and a $7680^3$-particle box with initial conditions from MDPL2, which we refer to as \texttt{GIZMO\_3k} and \texttt{GIZMO\_7k}, respectively. Unless stated otherwise, in this work we present results from the \texttt{GIZMO\_7k} run, but we have verified convergence using results from the \texttt{GIZMO\_3k}.

Particles inside the re-simulation regions were split into dark matter and gas particles with masses of $15.9\times 10^7\,h^{-1}\,{\rm M}_\odot$ and $2.95\times 10^7\,h^{-1}\,{\rm M}_\odot$, respectively, in the \texttt{GIZMO\_7k} run, and eight times larger in the \texttt{GIZMO\_3k} run. The Plummer-equivalent force softening used in the \texttt{GIZMO\_7k} run is $\epsilon_{\rm P}=2.5\,h^{-1}\,{\rm kpc}$, which corresponds to a spline softening length as used by \textsc{gadget-3} of $\epsilon_{\rm sp}=2.8\times\epsilon_{\rm P}=7\,h^{-1}\,{\rm kpc}$. When fitting radial profiles, we adopt a minimum radius equal to the halo convergence radius, which varies from halo to halo, but is always larger than $\epsilon_{\rm P}$ \citep{Power2003,Ludlow2019}.

The cosmological model for all runs is that of MDPL2, namely a flat $\Lambda$CDM universe with parameters from the Planck 2015 mission \citep{Planck2016}. The density parameters for the total matter, baryonic matter, and cosmological constant are $\Omega_{\rm m}=0.307$, $\Omega_{\rm b}=0.048$, and $\Omega_\Lambda=0.693$, respectively. The Hubble parameter is $h=0.678$, the \textsc{rms} density fluctuation on a scale of 8 Mpc is $\sigma_8=0.823$, and the spectral index of the primordial power spectrum is $n_{\rm s}=0.96$. 

Data are saved in 129 snapshots from $z=16.98$ to $z=0$. For each cluster, we use group catalogues constructed with the {\small AHF} halo finder \citep[cf.][]{Gill2004,Knollmann2009}, which includes information about the stellar and gas content of the main halo and its substructures.

\section{Methods}
\label{sec:methods}

In this section we outline the algorithm used to decompose the clusters into their orbiting and infalling components. We then go over the calculation and fitting of the density profiles.


\subsection{Orbiting-infalling decomposition}
\label{sec:orbit_decomposition}

\begin{figure*}
    \centering
    \includegraphics[width=0.87\linewidth]{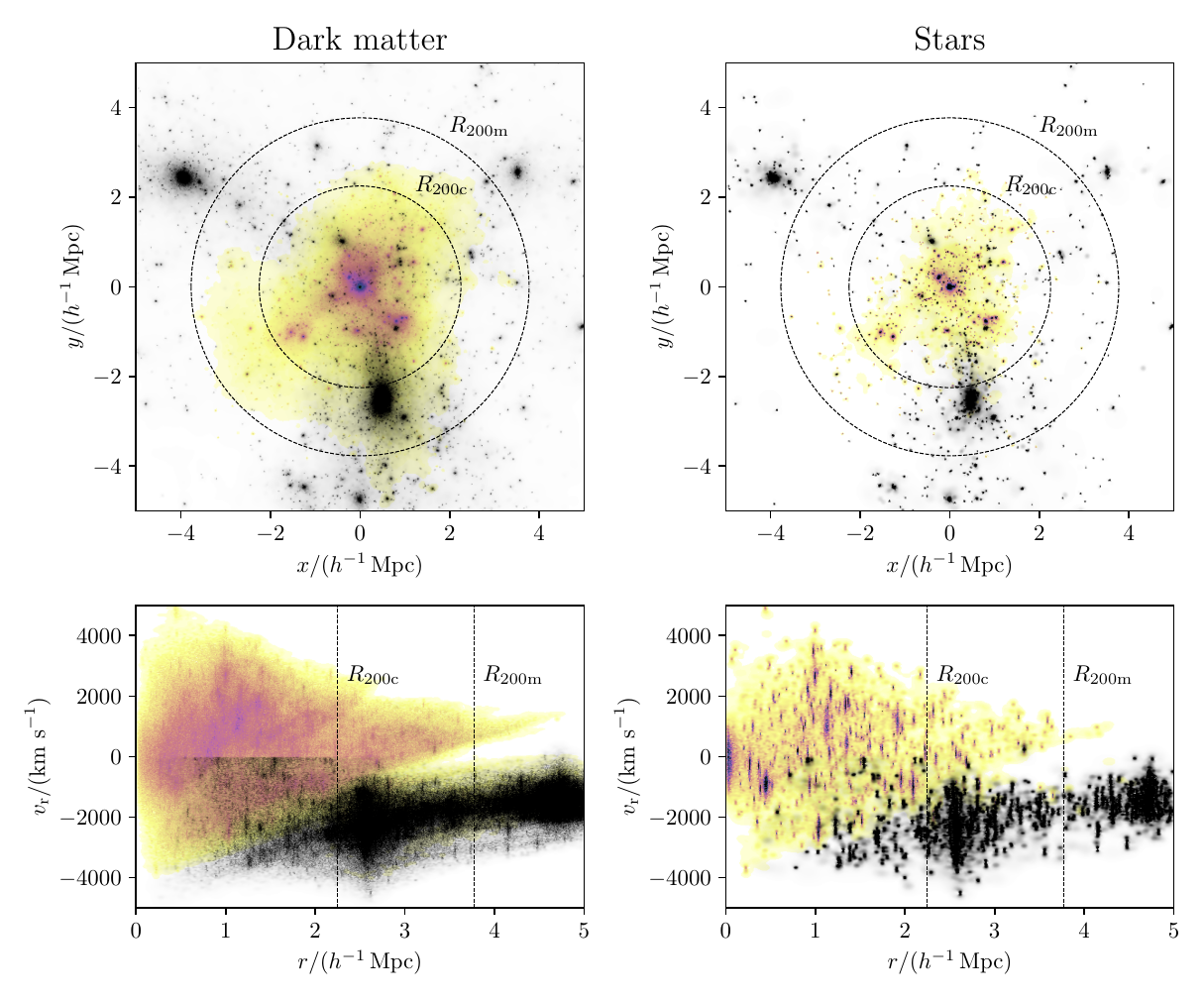}
    \caption{Cluster 001 from the \texttt{GIZMO\_3k} run, decomposed into orbiting (yellow-purple) and infalling (black) particles. The top row shows the $x$--$y$ projection of the dark matter (left) and stars (right), while the bottom row shows their radial phase space. The cluster's $R_{200{\rm c}}$ and $R_{200{\rm m}}$ are marked.}
    \label{fig:orbit-decomposition}
\end{figure*}
\begin{figure*}
    \centering
    \includegraphics[width=0.87\linewidth]{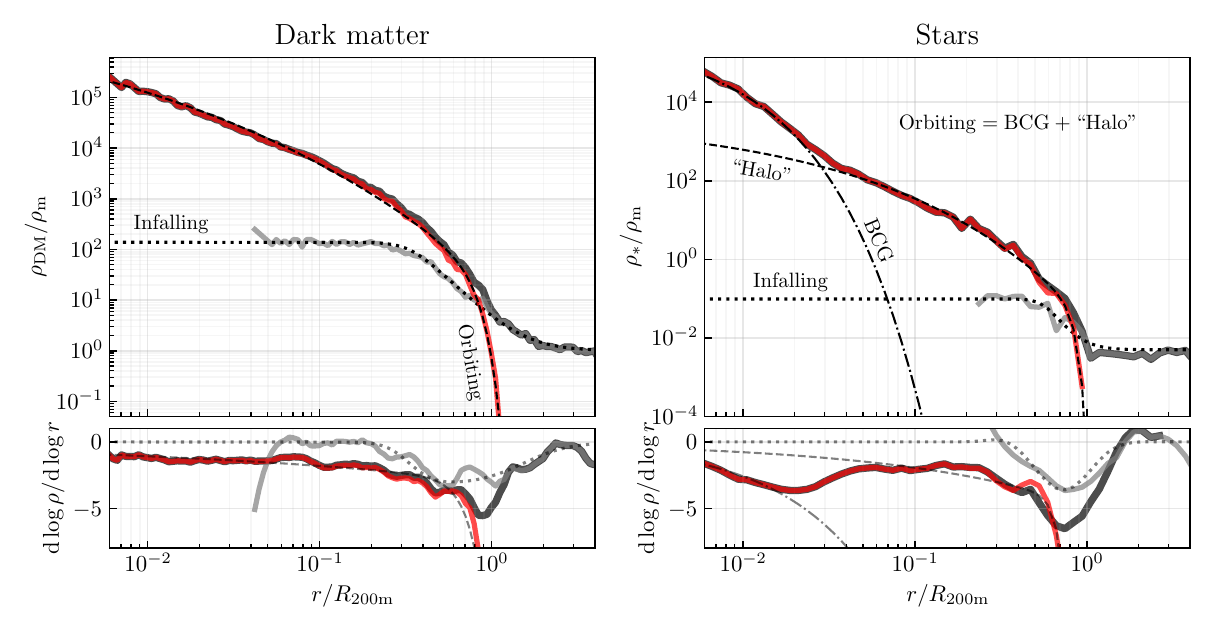}
    \caption{The density profiles of the orbiting and infalling components, fit using Equations \ref{eq:orbiting-fitting-function-dm}--\ref{eq:infalling-fitting-function}. The BCG component is confined to the inner regions, while the stellar ``halo'' takes on a similar shape to the orbiting profile of the dark matter. The logarithmic slope profiles are shown in the bottom panels.}
    \label{fig:decomposed-profiles}
\end{figure*}

The cluster center of mass and velocity center are calculated using particles of all types (dark matter, stars, and gas) that are contained within $R_{200{\rm c}}$ of the position of peak density determined by \textsc{ahf}. To identify the splashback of stars and dark matter in the simulated clusters, we classify all particles as ``orbiting'' or ``infalling'' depending on if the particle has passed through at least one pericenter about the center of mass of the cluster. This is done by recording a change in the sign of a particle's physical radial velocity (peculiar plus Hubble flow) from negative to positive. To filter out radial velocity sign flips due to orbits within subhalos, we also require that particles have traversed an angle of at least $\pi/2$ around the halo center since entering the cluster region. Any particle that has not had such a sign change is regarded as ``infalling''. Though this classification technically considers positive radial velocity particles beyond the turn-around radius as ``infalling'', we limit our analysis of the density profiles to radii safely within this boundary. We also do not track gas particles, meaning any stars that form in orbit are temporarily classified as ``infalling'' until their first pericenter.

We identify cluster progenitors by selecting the 100 most-bound particles of each halo and identifying the halo in the previous snapshot that contains the highest fraction of them. This gives us a halo position $\boldsymbol{X}(z)$ and velocity $\boldsymbol{V}(z)$ at every snapshot relative to which we calculate the radial velocity of every particle. That is, for a particle with comoving position $\boldsymbol{x}$ and peculiar velocity $\boldsymbol{v}=a\dot{\boldsymbol{x}}$ (where $a$ is the scale factor) we calculate $v_r=[(\boldsymbol{v}-\boldsymbol{V})+H\boldsymbol{r}]\cdot\hat{\boldsymbol{r}}$, where $\boldsymbol{r}=a(\boldsymbol{x}-\boldsymbol{X})$.

The result of the orbiting-infalling decomposition for the most massive cluster in the \texttt{GIZMO\_3k} run is shown in Figure \ref{fig:orbit-decomposition}. The top row shows the $x$--$y$ projection of the dark matter (left) and stars (right), while the bottom row is their radial phase space (with Hubble flow added). The phase space shows a clear separation between the infalling stream and the orbiting material, with the infalling material abruptly transitioning into orbiting material at the point of pericenter, $v_{\rm r}=0$. The orbiting particles at the largest radii still have some positive radial velocity, meaning they are still moving outwards and have not quite ``splashed back''. Visually, the distribution of the dark matter is well traced by the stellar distribution, and both the orbiting dark matter and orbiting stars extend significantly past $R_{200{\rm c}}$ and out to about $R_{200{\rm m}}$. See e.g. \citet{Diemer2017,Bakels2021} for discussion on the extent of the orbits of halo material.


\subsection{Cluster density profiles}
\label{sec:density_profiles}

The density profiles of the clusters are calculated at $z=0$ in logarithmically-spaced bins between $0.01\,{\rm Mpc}$ and $15\,{\rm Mpc}$ from the position of peak density, as determined by \textsc{ahf}. We further bin the radial shells into segments of equal solid angle and compute the median, which suppresses local fluctuations in density due to anisotropic substructure. For the lower-resolution \texttt{GIZMO\_3k} run, we choose 128 radial bins and 48 angular bins for the dark matter (average of $\sim$4,000 particles per radial bin), and 64 radial bins and 12 angular bins for the stars (average of $\sim$600 particles per radial bin). For the high-resolution \texttt{GIZMO\_7k} run, we choose 128 radial bins and 48 angular bins for both the dark matter and the stars (average of $\sim$30,000 and $\sim$3,000 particles per radial bin, respectively). As well as the total density profile, we also calculate the density profiles of the orbiting and infalling material using the same binning. To calculate the logarithmic density slope profiles, we take the derivative using a $4^{\rm th}$-order Savitzsky-Golay filter over the 20 nearest radial bins for the 128 bin profiles, and the nearest 10 radial bins for the 64 bin profiles. The profiles are calculated at $z=0$ using all particles of the relevant type, whether gravitationally bound or not.

The dark matter and stellar density profile of the most massive cluster in the \texttt{GIZMO\_3k} run (cluster 001) is shown in Figure \ref{fig:decomposed-profiles}. The profile of the orbiting material is shown in red, while that of the infalling material is in grey. The bottom panels show the corresponding logarithmic slope profiles. The radius is normalized by $R_{200{\rm m}}$, the radius enclosing an average density equal to $200$ times the mean matter density of the universe, $\rho_{\rm m}$. All overdensity radii and masses we use in this work are calculated using the total density (stars, gas, and dark matter). The outer profiles of the dark matter and stars ($\gtrsim 0.1\times R_{200{\rm m}}$) are---at least qualitatively---very similar in shape (though very different in magnitude): both feature a rapidly steepening orbiting profile and an infalling profile that decreases much more gradually with radius, eventually overtaking the orbiting density at $\sim R_{200{\rm m}}$ before plateauing to the background density. In the inner regions the profiles are very different, with the stellar profiles having an enhancement in the density due to the cluster's central galaxy (the ``brightest cluster galaxy'', or BCG). The logarithmic slope of the total profiles exhibit the well-known minimum near $R_{200{\rm m}}$, which occur at the same radius for the dark matter and stars and coincide with the truncation in the slope of the orbiting profile.


\subsection{Profile fitting functions}
\label{sec:fitting-functions}

To understand how the stellar orbiting and infalling profiles quantitatively relate to those of the dark matter we use the fitting functions introduced by \citet{Diemer2023}, which have been shown to recover the orbiting and infalling profiles of the dark matter from the total profile reasonably well \citep{Diemer2025}. The fitting function for the orbiting profile is an Einasto profile that is exponentially suppressed by a truncation term:
\begin{equation}
\rho_{\rm orb, DM}(r)=\rho_{\rm s}e^{S(r)},\notag
\end{equation}
where
\begin{equation}
\label{eq:orbiting-fitting-function-dm}
S(r)=-\frac{2}{\alpha}\left[\left(\frac{r}{r_{\rm s}}\right)^\alpha-1\right]-\frac{1}{\beta}\left[\left(\frac{r}{r_{\rm t}}\right)^\beta-\left(\frac{r_{\rm s}}{r_{\rm t}}\right)^\beta\right].
\end{equation}
Here $r_{\rm s}$ and $\rho_{\rm s}=\rho(r_{\rm s})$ are the scale radius and scale density, $\alpha$ is the exponent of the power-law inner slope, and $r_{\rm t}$ and $\beta$ are the radius and steepness of the orbiting profile truncation.

Though this function describes the orbiting dark matter well, it is not well-suited to fitting the inner part of the stellar profile where the BCG becomes important. We therefore add a S\'ersic ``BCG'' term to Equation \ref{eq:orbiting-fitting-function-dm}. We refer to the part described by Equation \ref{eq:orbiting-fitting-function-dm} as the stellar ``halo'', so that the orbiting fitting function for the stars is given by
\begin{align}
\label{eq:orbiting-fitting-function-stars}
\rho_{\rm orb,\ast}(r)&=\rho_{\rm halo}(r)+\rho_{\rm BCG}(r)\notag\\
&=\rho_{\rm s}e^{S(r)}+\rho_0\exp\left\{-c\left[\left(\frac{r}{r_0}\right)^k-1\right]\right\}
\end{align}
where $c$ and $k$ control the shape of the BCG profile and we refer to $r_0$ and $\rho_0=\rho(r_0)$ as the BCG radius and characteristic density, respectively. We do not assume any relationship between $c$ and $k$ and instead allow them to vary freely when fitting. We also emphasise that although $\rho_{\rm orb,DM}$ and $\rho_{\rm halo}$ have the same functional form, we refer to them with different names because they represent different quantities---namely the dark matter orbiting profile and the ``halo'' part of the stellar orbiting profile.

The infalling model is given by the following function that flattens out to fixed densities at small and large radii:
\begin{align}
\label{eq:infalling-fitting-function}
\rho_{\rm inf}(r)=\rho_\infty\left(\frac{\delta_1}{\sqrt{(\delta_1/\delta_{\rm max})^2+(r/R_{200{\rm m}})^{2s}}}+1\right),
\end{align}
where $\delta_1$ is the normalization, $s$ controls the slope, and $\rho_\infty$ is the density at large radii. Although the statistical contribution from large-scale structure begins to dominate the density profile at large radii, we only fit out to a few times $R_{200{\rm m}}$ and so we do not encounter this transition. Since the shape of the stellar infalling profile is similar to that of the dark matter, we expect it to be well-fit by Equation \ref{eq:infalling-fitting-function}.

We fit the profiles using least-squares minimization in the logarithm of the density. The radial bins are weighted equally and the fitting is done between $4\,R_{200{\rm m}}$ and the convergence radius predicted by \citet{Power2003} and \citet{Ludlow2019} for the dark matter-only case. The results of fitting Equations \ref{eq:orbiting-fitting-function-dm}--\ref{eq:infalling-fitting-function} to the orbiting and infalling profiles of an individual cluster (cluster 001) is shown in Figure \ref{fig:decomposed-profiles}.

As well as fitting the orbiting and infalling profiles of the individual clusters, we also fit profiles averaged in bins of accretion rate to investigate any possible systematic differences. We define the accretion rate at $z=0$ as
\begin{align}
\label{eq:accretion-rate}
\Gamma\equiv\Gamma_{200{\rm m}}=\frac{\Delta\log\,M_{200{\rm m}}}{\Delta\log\,a},
\end{align}
where $a$ is the scale factor and $\Delta$ represents a change over one dynamical time, defined as the crossing time $2\,R_{200{\rm m}}/V_{200{\rm m}}$ $\propto (G\rho_{\rm m})^{-1/2}$. At $z=0$, one dynamical time in the past corresponds to $a\approx 0.769$. Here, as with everywhere else in this work, $M_{200{\rm m}}$ is calculated using the total density (stars, gas, and dark matter). Progenitor halos are identified as those that contain the largest fraction of the most-bound particles of the $z=0$ halos. We choose six equal-width bins between $\Gamma=0$ and $\Gamma=6$ and calculate the median logarithmic density profile in each bin, in units of $r/R_{200{\rm m}}$. The first four bins contain between 51 and 102 clusters, while the remaining two contain 17 and 10.

\section{Results}
\label{sec:results}

\begin{figure}
    \centering
    \includegraphics[width=0.8\linewidth]{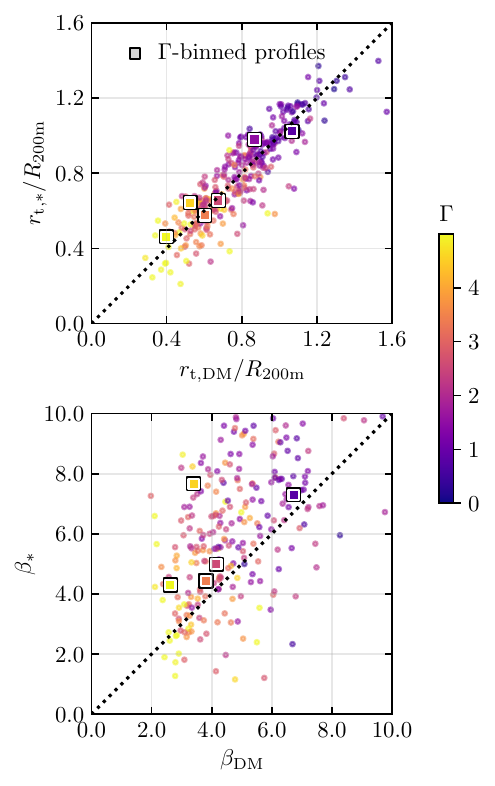}
    \caption{Comparison between $r_{\rm t}$ and $\beta$ obtained from fitting the dark matter and stellar orbiting profiles. The points are colored by accretion rate (Equation \ref{eq:accretion-rate}). The square points are the result of fitting the $\Gamma$-binned median profiles.}
    \label{fig:rt-beta-stars-vs-dm}
\end{figure}

In this section, we compare the fit parameters from the stellar and dark matter profiles, focusing specifically on the parameters $r_{\rm t}$ and $\beta$ that describe the splashback region. The first subsection shows results from fitting the 3D orbiting and infalling profiles, and so represent the ``true'' parameters. In the subsequent subsection, we present results from fitting the projected total profiles to determine how well the truncation parameters can be recovered from observed profiles. We leave the comparison of the infalling component to Section \ref{sec:infalling-profile} and discussion of the inner profile to Section \ref{sec:inner-profile}.

\subsection{Fit parameter relations}

Figure \ref{fig:rt-beta-stars-vs-dm} shows how $r_{\rm t}$ (left) and $\beta$ (right) compare between dark matter and stars. $r_{\rm t}$ follows a one-to-one relation with relatively small scatter---in other words, the splashback radius of the stars and dark matter in clusters is the same on average. This confirms what was strongly hinted at by the \citet{Deason2021} results on the radius of steepest slope. On the other hand, $\beta$ does not follow a one-to-one relation, with the truncation in the stars generally being steeper than that of the dark matter. This is also consistent with the \citet{Deason2021} finding that the steepest slope is steeper for the stars than the dark matter. The points are colored by accretion rate (Equation \ref{eq:accretion-rate}), which reveals a clear gradient in $r_{\rm t}$ and a less clear but still noticeable gradient in $\beta$, with each increasing with decreasing $\Gamma$. That is, the splashback boundary is sharper and is found at a larger fraction of $R_{200{\rm m}}$ for slowly accreting clusters. The cyan squares show the $r_{\rm t}$ and $\beta$ from fitting to the median of the $\Gamma$-binned profiles, and they are consistent with the individual measurements.

Figure \ref{fig:rt-beta-stars-vs-dm} shows that the outer orbiting profile---i.e. the splashback region---is very similar between dark matter and stars, albeit with the stellar density dropping off more steeply on average. We should therefore expect the splashback region of the stars in clusters to be an accurate tracer for the splashback region of the dark matter.

\begin{figure}
    \centering
    \includegraphics[width=1.\linewidth]{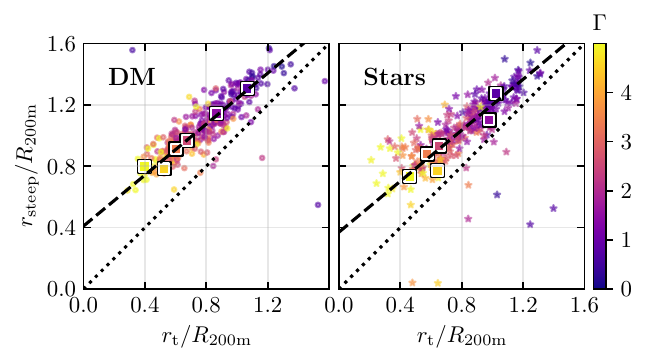}
    \caption{Comparison between the radius of steepest slope, $r_{\rm steep}$ and $r_{\rm t}$ for the dark matter and stars. The points are colored by accretion rate (Equation \ref{eq:accretion-rate}). The square points are the result of fitting the $\Gamma$-binned median profiles. The black dashed lines are linear fits given by $r_{\rm steep}/R_{200{\rm m}}=0.83(r_{\rm t}/R_{200{\rm m}})+0.41$ for the dark matter and $r_{\rm steep}/R_{200{\rm m}}=0.83(r_{\rm t}/R_{200{\rm m}})+0.37$ for the stars.}
    \label{fig:rsteep-vs-rt}
\end{figure}

\begin{figure}
    \centering
    \includegraphics[width=1.\linewidth]{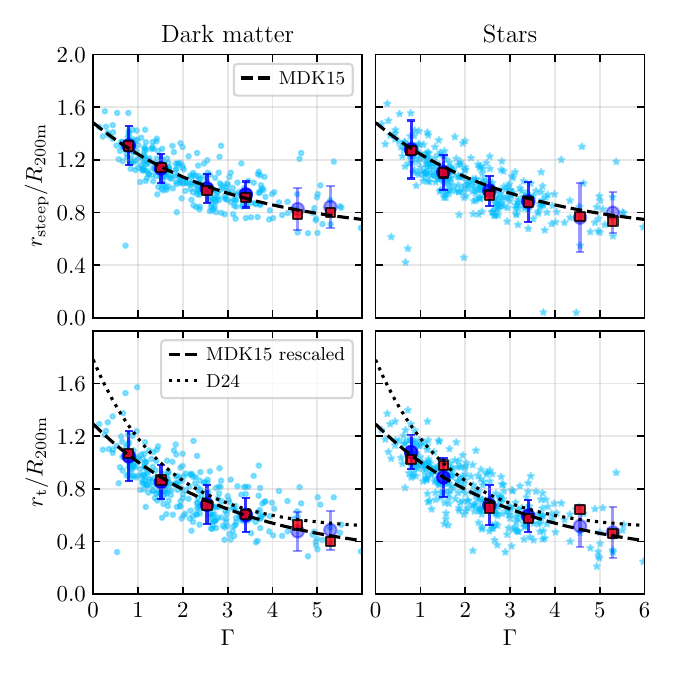}
    \caption{$r_{\rm steep}$ and $r_{\rm t}$ as a function of accretion rate (Equation \ref{eq:accretion-rate}). The light blue points are measurements for individual profiles, while the dark blue points are the median and standard deviation in equally-spaced bins in $\Gamma$. The red squares are from fitting the $\Gamma$-binned median profiles. The black dashed line in the top row is the relation from \citet{More2015}, while in the bottom row it is the same relation rescaled using the best-fit $r_{\rm t}$--$r_{\rm steep}$ relation shown in Figure \ref{fig:rsteep-vs-rt}. The black dotted line is the relationship from \citet{Diemer2025}.}
    \label{fig:rt-vs-accretion}
\end{figure}

\begin{figure*}
    \centering
    \includegraphics[width=0.9\linewidth]{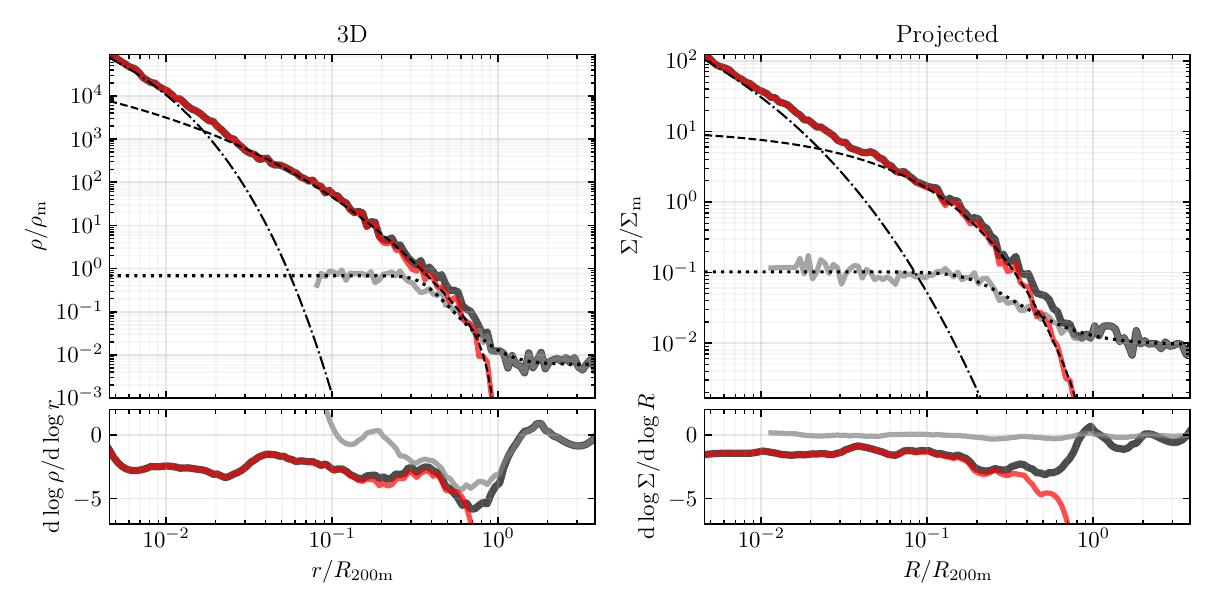}
    \caption{The decomposed stellar density profile of the same cluster from the \texttt{GIZMO\_7k} run (cluster 005) in 3D (left) and in projection (right), along with the best-fit curves. The projected density has been normalized by the mean matter density multiplied by 30 Mpc, the diameter of the cluster region.}
    \label{fig:3D-projection-comparison}
\end{figure*}

The truncation radius is a more direct measure of the size of the splashback boundary (technically the orbiting boundary, since the outermost particles may not have reached apocenter yet) than the radius of steepest slope, which is determined in part by the infalling material. In perfectly spherical collapse, the truncation of the orbiting profile is infinitely sharp and the two radii coincide. However, in realistic simulations, mass is accreted anisotropically and particle orbits can take on a variety of shapes, which has the effect of smoothing out the edge of the orbiting profile. In this case, the radius of steepest slope is not simply related to the edge of the orbiting material, but is rather the point at which the infalling profile becomes dominant enough to strongly influence the total slope. In Figure \ref{fig:rsteep-vs-rt} we plot $r_{\rm steep}$ measured from the total profiles against $r_{\rm t}$. Again, the points are coloured by $\Gamma$ and show the gradient expected from the now well-known $r_{\rm steep}$--$\Gamma$ relation. The cyan points are measurements of $r_{\rm steep}$ from the $\Gamma$-binned median total profile. The $r_{\rm t}$--$r_{\rm steep}$ relation seems to be described well by a line with slope of $\sim 0.83$ and $r_{\rm steep}(r_{\rm t}=0)/R_{200{\rm m}}\approx 0.41$ for the dark matter and $0.37$ for the stars (black dashed lines). On average, the radius of steepest slope is $\sim 0.25\,R_{200{\rm m}}$ larger than the truncation radius.

The gradient with accretion rate seen in the radii in Figures \ref{fig:rt-beta-stars-vs-dm} and \ref{fig:rsteep-vs-rt} reflect the now well-established relationship between the splashback radius and accretion rate. This relationship was first reported by \citet{Diemer2014} for the radius of steepest slope in dark matter-only cosmological simulations, while its theoretical basis was established by \citet{Adhikari2014} using a model based on spherical collapse. The relation was shown by \citet{More2015} to be well-described by an exponential of the form $r_{\rm steep}/R_{200{\rm m}}=a+be^{-\Gamma/c}$, and this same function was shown by \citet{Deason2021} to hold for the stellar $r_{\rm steep}$ of 34 C-EAGLE clusters. Recently, \citet{Diemer2025} showed that the relationship between $r_{\rm t}$ and $\Gamma$ for stacked halos is also well-described by an exponential that has remarkably little dependence on halo mass or cosmology.

\begin{figure}
    \centering
    \includegraphics[width=1.\linewidth]{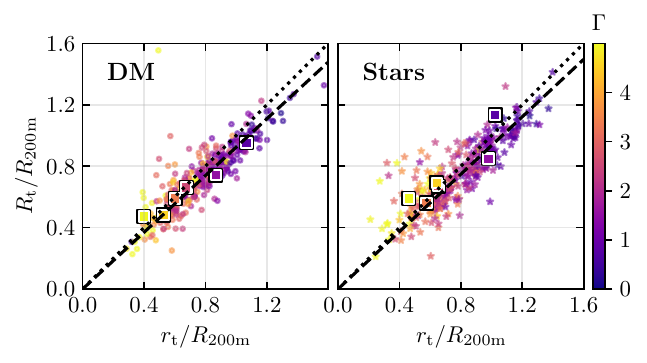}
    \includegraphics[width=1.\linewidth]{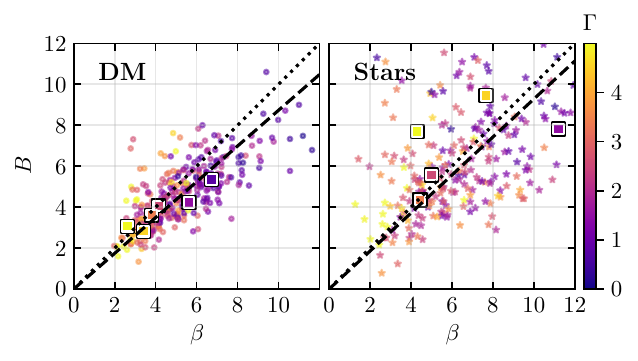}
    \caption{Comparison between 3D and projected $r_{\rm t}$ and $\beta$. The points are colored by accretion rate (Equation \ref{eq:accretion-rate}). The square points are the result of fitting the $\Gamma$-binned median profiles. The black dashed line is a linear fit.}
    \label{fig:rt-beta-projected-vs-3d}
\end{figure}

In Figure \ref{fig:rt-vs-accretion} we plot the accretion rate dependence of the stellar $r_{\rm t}$ and $r_{\rm steep}$. For comparison, we also show the corresponding plots for the dark matter. The light blue points are individual clusters, while the dark blue points show the median in the same accretion rate bins as we use to calculate median profiles. The measurements from the median profiles are shown in red squares. The increased transparency on the right-most points indicates that they correspond to bins with $<20$ clusters. The fitting function for $\Gamma_{200{\rm m}}$--$r_{\rm steep}$ from \citet{More2015} is plotted on the top row and agrees well with our data. The fitting function for $\Gamma_{200{\rm m}}$--$r_{\rm t}$ from \citet{Diemer2025} is plotted on the bottom row and agrees at large $\Gamma_{200{\rm m}}$, but rises more rapidly than our data towards the small end. For comparison, we plot the \citet{More2015} relation rescaled according to the $r_{\rm t}$--$r_{\rm steep}$ relation we find from Figure \ref{fig:rsteep-vs-rt}, and it seems to be a better fit.

Comparing the dark matter and the stars, we see that they follow the same relationship with $\Gamma_{200{\rm m}}$ and that the scatter is very similar. The accretion rate can therefore be estimated from measurements of the stellar $r_{\rm t}$---assuming one has access to the 3D orbiting profile of the stars. However, observations only have access to the total stellar density, and in projection at that. We therefore now turn to the projected total profiles to assess how these results can be applied to observations.


\subsection{Recovering $r_{\rm t}$ from observable profiles}

So far, we have shown results from fitting the three-dimensional orbiting and infalling profiles. However, real clusters are observed in projection and information about the orbiting and infalling material is not directly accessible. In this section, we show results from fitting the \textit{total}, \textit{projected} density profiles of the stars, as well as those of the dark matter for comparison. This means using no information about the orbiting and infalling identification and instead fitting the density profile with the total model, which in projection we write as $\Sigma(r)=\Sigma_{\rm orb}(r)+\Sigma_{\rm inf}(r)$.

We calculate the projected density profiles using the same radial binning as we used for the 3D profiles. The angular binning is also done with the same number of bins, but this time in equal angle rather than equal solid angle. We project the entire cluster region along the $x$, $y$ and $z$ axes, and take the mean of the resulting profiles. Since the zoom regions contain only the cluster and its correlated surroundings, there is no contribution to the profile from uncorrelated field galaxies. We therefore do not apply any background subtraction (see \citealt{Sun.etal.2025} for the effect of the projection window depth on the splashback feature). The calculation of the density slope profiles is the same as for the 3D case. For clarity, we refer to the projected density with the symbol $\Sigma$ and other quantities with upper case, e.g. $R_{\rm t}$ and $B$ for the truncation radius and steepness.

In projection, the shape of the density profile is modified. For the stars, the splashback region and BCG-stellar halo transition are smoothed out. This can be seen in Figure \ref{fig:3D-projection-comparison}, which compares the 3D and projected stellar profile of one of the clusters in the \texttt{GIZMO\_7k} run. Despite this, the profiles are still well-fit by Equations \ref{eq:orbiting-fitting-function-dm}--\ref{eq:infalling-fitting-function}, generalizing earlier work on the similarity between 3D and projected total profiles (e.g. \citealt{Merritt2005}). It is reasonable to expect that the parameters of the 3D and projected profiles should be related. Figure \ref{fig:rt-beta-projected-vs-3d} compares the truncation radius and truncation steepness from fitting the 3D and projected orbiting profiles (\textit{not} the total profiles), for the dark matter (left) and stars (right). The projected $R_{\rm t}$ is slightly smaller than the 3D $r_{\rm t}$ by a factor of $\sim 0.92$ for the dark matter and $0.94$ for the stars (black dashed lines). This is similar to the factor of $0.9$ found by \citet{Deason2021} when comparing the 3D and projected steepest slope radii. The projected $B$ of the dark matter is also smaller than its 3D counterpart by a factor of $\sim 0.87$. This may also be the case for the stars, but the scatter is significantly larger.

\begin{figure}
    \centering
    \includegraphics[width=1.0\linewidth]{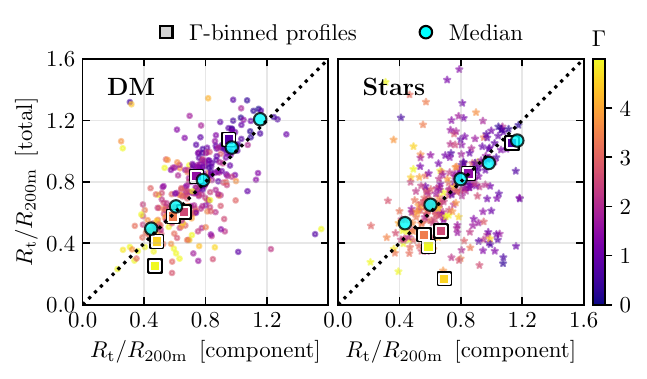}
    \caption{Comparison between the projected truncation radius obtained by fitting the orbiting profile (labelled component) and the total profile. The points are colored by accretion rate (Equation \ref{eq:accretion-rate}). The cyan points are the median total-profile $R_{\rm t}$ calculated in five equally-spaced bins. The square points are the result of fitting the $\Gamma$-binned median profiles.}
    \label{fig:rt-component-vs-total}
\end{figure}

We have chosen to show the results from fitting the orbiting profiles in Figure \ref{fig:rt-beta-projected-vs-3d} to see the ``true'' effect of projection on the outer profile. Fitting the total profile, as is required when information about the orbiting and infalling material is inaccessible, inevitably sacrifices some accuracy as the model attempts to recover the ``best-fitting'' orbiting and infalling profiles. As shown by \citet{Diemer2025}, the model is able to do this relatively successfully---at least for binned profiles. In Figure \ref{fig:rt-component-vs-total}, we compare the $R_{\rm t}$ obtained from fitting the orbiting profile (which we refer to as ``component'') to that obtained from fitting the total profile, where both are in projection. We find that fitting the total profiles introduces a scatter of $\sim 0.3$ around the ``true'' $R_{\rm t}$, for both the dark matter and the stars. The median total $R_{\rm t}$, calculated in five bins in the component $R_{\rm t}$ and denoted by the red circles, shows that there is no significant $R_{\rm t}$-dependent systematic error introduced by fitting the total profiles of the dark matter, while there is possibly a slight systematic error for the stars.

\begin{figure}
    \centering
    \includegraphics[width=1.0\linewidth]{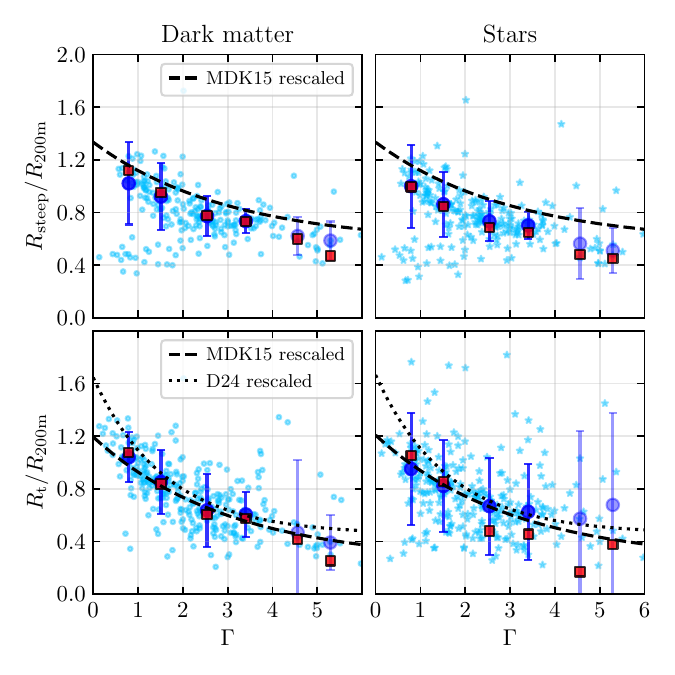}
    \caption{$R_{\rm steep}$ and $R_{\rm t}$ as a function of accretion rate (Equation \ref{eq:accretion-rate}). The light blue points are measurements for individual profiles, while the dark blue points are the median and standard deviation in equally-spaced bins in $\Gamma$. The red squares are from fitting the $\Gamma$-binned median profiles. The black dashed line in the top row is the relation from \citet{More2015}, rescaled using the factor of $r_{\rm steep}/R_{\rm steep}=0.9$ found by \citet{Deason2021}. The black dashed in the bottom row is the same curve further rescaled using the best-fit $r_{\rm t}$--$r_{\rm steep}$ relation shown in Figure \ref{fig:rsteep-vs-rt}. The black dotted line is the relationship from \citet{Diemer2025}, rescaled by the slope of the best-fit line shown in Figure \ref{fig:rt-beta-projected-vs-3d}.}
    \label{fig:rt-vs-accretion-total-projected}
\end{figure}

Figure \ref{fig:rt-vs-accretion-total-projected} is the equivalent of Figure \ref{fig:rt-vs-accretion} for the $R_{\rm t}$ and $R_{\rm steep}$ measured from the projected total profiles. This time, the \citet{More2015} relation for $r_{\rm steep}$ has been multiplied by the $r_{\rm steep}/R_{\rm steep}$ factor found by \citet{Deason2021}, while the $r_{\rm t}$ relations have been multiplied by the $r_{\rm t}/R_{\rm t}$ factor we found from Figure \ref{fig:rt-beta-projected-vs-3d}.

The scatter in both $R_{\rm steep}$ and $R_{\rm t}$ is larger compared to the 3D and 3D component measurements, however this increase is more noticeable for $R_{\rm t}$ due to the error introduced when fitting to the total profiles (see Figure \ref{fig:rt-component-vs-total}). Despite this, the accretion rate dependence is still discernable.

The $R_{\rm steep}$--$\Gamma$ plots feature a small group of points that lie below the relation at low $\Gamma$. These are cases where the $2^{\rm nd}$ caustic---an additional, usually shallower caustic at smaller radius---actually becomes steeper than the $1^{\rm st}$ caustic when viewed in projection. The measurement of $R_{\rm t}$ seems to be unaffected by this, suggesting the fitting method may have an advantage in these cases.

\subsection{The infalling profile}\label{sec:infalling-profile}

\begin{figure*}
    \centering
    \includegraphics[width=0.9\linewidth]{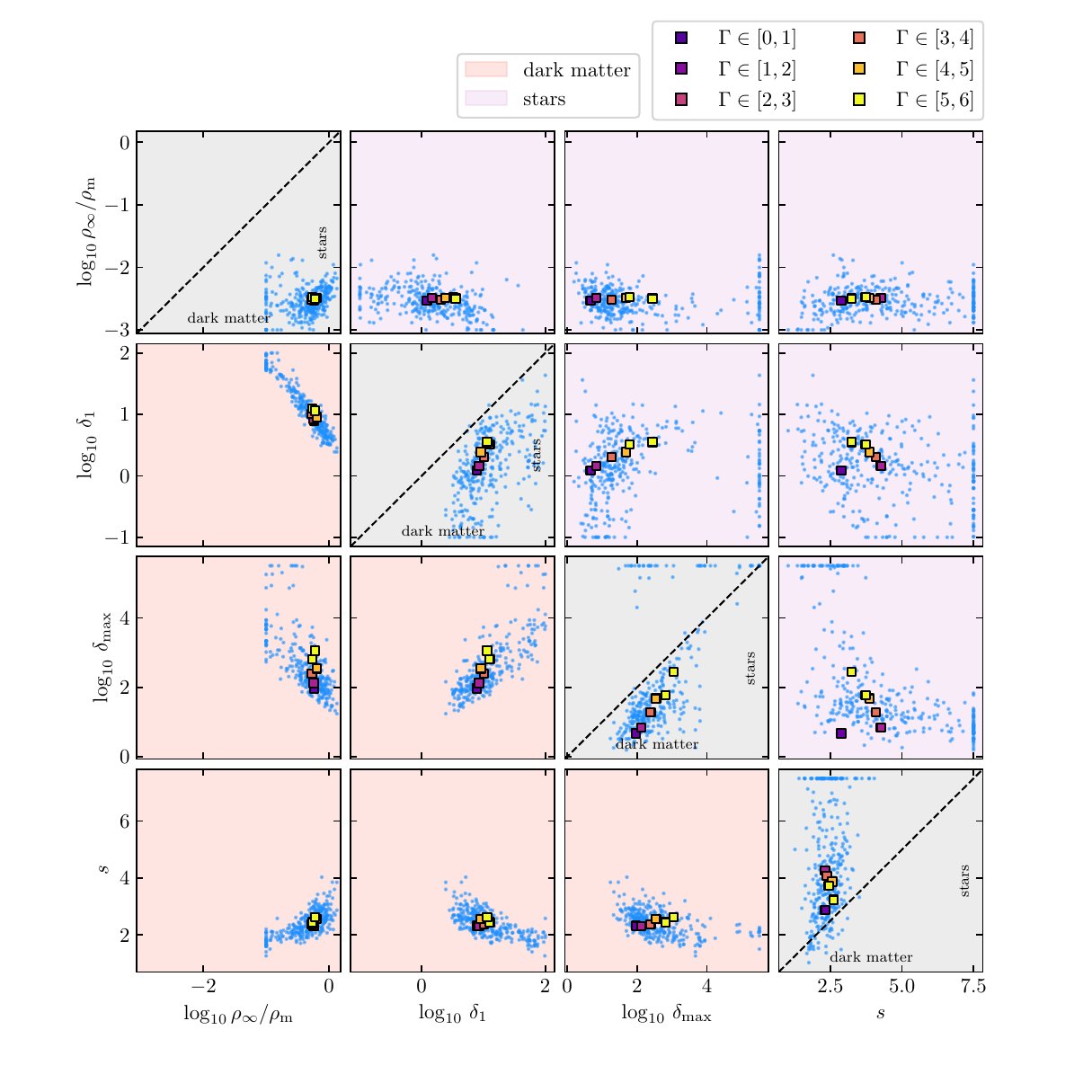}
    \caption{Best-fit parameters for the 3D infalling profiles. The bottom-left (red panels) and top-right (purple panels) show the correlations between each parameter for the dark matter and stars, respectively. The gray diagonal panels compare the parameters between the stars and dark matter. The square points are the result of fitting the median of profiles binned by $\Gamma$, as specified in the legend.}
    \label{fig:infalling-params}
\end{figure*}

\begin{figure*}
\centering
\includegraphics[width=0.4\linewidth]{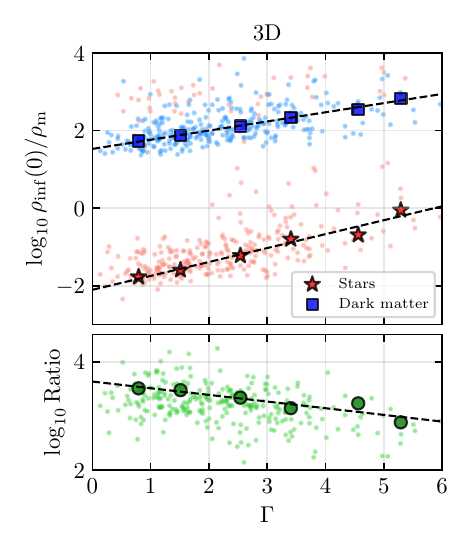}
\includegraphics[width=0.4\linewidth]{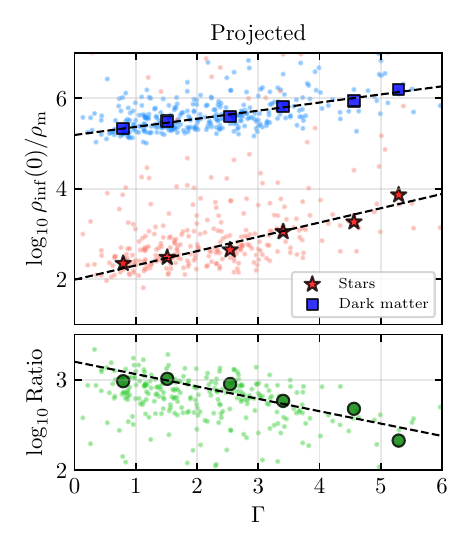}
\caption{Best-fit infalling density at $r=0$ for dark matter (blue) and stars (red) in 3D and projection. The larger, dark points are from the median $\Gamma$-binned profiles. The bottom panels show the ratio between the blue and red points. The black dashed lines are lines of best fit to the large points.}
\label{fig:central-infalling-density-vs-accretion-rate}
\end{figure*}

So far we have focused on the outer orbiting profile of the stars and dark matter. In this section we turn to the parameters of the infalling profile---the profile of material that has yet to make its first pericentric passage---to see how they compare between the dark matter and stars.

Figure \ref{fig:infalling-params} shows the distribution of the parameters from fitting the 3D infalling profiles. The bottom-left (red panels) and top-right (purple panels) show the correlations between each parameter for the dark matter and stars, respectively. The gray diagonal panels compare the parameters between the stars and dark matter. The square points are the result of fitting the median $\Gamma$-binned profiles.

Unsurprisingly, the asymptotic density $\rho_\infty/\rho_{\rm m}$ is much smaller for the stars, as seen in the top-left plot. The $\Gamma$-binned points all lie on top of one another, suggesting that $\rho_{\infty}/\rho_{\rm m}$ does not evolve with $\Gamma_{200{\rm m}}$. This makes sense since material far outside $R_{200{\rm m}}$ has not yet been accreted. By taking the mean of the median binned points, we find that the asymptotic dark matter density is $\rho_\infty^{\rm DM}/\rho_\infty^\ast\sim 182$ times that of the stars.

The normalization of the stellar infalling profile, $\log_{10}\delta_1$, is smaller than its dark matter counterpart and increases with it, though not as a one-to-one relationship. The case is similar for $\log_{10}\delta_{\rm max}$, though the slope is closer to 1.

The infalling slope, $s$, is generally larger for the stars but has significantly larger scatter than the dark matter.

Based on the $\Gamma$-binned points, $\delta_1$, $\delta_{\rm max}$, and $s$ all appear to evolve with accretion rate. However, plotting these against accretion rate shows that, for $\delta_1$ and $s$, this is mostly scatter. $\delta_{\rm max}$, on the other hand, evolves unambiguously with accretion rate. The scatter in this relation is reduced when we instead consider the predicted infalling density at the center of the halo: $\rho_{\rm inf}(0)=\rho_\infty(\delta_{\rm max}+1)$. This is shown in Figure \ref{fig:central-infalling-density-vs-accretion-rate}. The dependence on $\Gamma_{200{\rm m}}$ for both the stars (red) and dark matter (blue) is linear with minimal scatter. The slope of the stellar relation is slightly larger than that of the dark matter, both in 3D and projection. The ratio of the dark matter and stellar density is shown in green in the sub-plots.


\subsection{The inner profile}\label{sec:inner-profile}

In this section we turn to the inner density profile of the stars and discuss its relation to the dark matter.

In our model, the orbiting density profile of the stars is composed of a S\'ersic BCG and a stellar ``halo'' fit with the truncated Einasto profile of \citet{Diemer2025} (see Section \ref{sec:fitting-functions}). As with the unmodified Einasto profile, the shape of the inner profile is controlled by the inner slope exponent $\alpha$ and the scale radius $r_{\rm s}$---the radius at which the logarithmic slope equals $-2$. Unlike for the outer profile parameters, we find that the inner stellar profile parameters do not track those of the dark matter well. The stellar $r_{\rm s}$ tends to underestimate that of the dark matter, while $\alpha$ has significant scatter around the one-to-one line. These comparisons are shown in Figure \ref{fig:rs-alpha-stars-vs-dm}.

At first glance, this discrepancy suggests that the dark matter and stellar profiles may not share a common scale radius, or that a well-defined scale radius may not exist for the stellar component at all. This would not be unexpected given the presence of the central BCG, and the growing importance of dynamical friction and tidal stripping toward the cluster centre, which reshape the stellar and dark matter distributions in different ways.

To test this, we identify $r_{\rm s}$ non-parametrically by locating the peak of the $\rho\,r^2$ profile. To do this, we take the median of profiles binned by $\Gamma$, considering both the angular-median profiles used throughout this work and conventional spherically averaged profiles. We further divide the stellar component into stars bound to halos (as identified by \textsc{ahf}) and the remaining diffuse BCG+ICL component. This is shown in Figure \ref{fig:DM-stellar-profile-comparison-angmean} for the spherically-averaged case. The stellar profiles, shown in yellow, have been rescaled by the median value of $1.5\times M_{\rm DM}(<R_{200{\rm c}})/M_\ast(<R_{200{\rm c}})$ in each bin so that they lie approximately on top of the dark matter curves (black). The density profiles of the stars in satellites have a clear peak that coincides with that of the dark matter, as shown by the logarithmic slope profiles in the bottom panel. Indeed, the satellite profile in the high-accretion-rate bin closely follows the dark matter profile until about $\sim 0.5 \, r_{\rm s}$ before falling off more rapidly towards smaller radii dominated by the BCG. In the low-accretion case, $r_{\rm s}/R_{200{\rm m}}$ is smaller (i.e. the concentration is larger) and the fall-off in the satellite density occurs close to $r_s$. The BCG also extends to larger $r/R_{200{\rm m}}$, which conceals the peak at these low accretion rates.

These results suggest that, at high accretion rates, the total stellar profile accurately traces the dark matter profile around the scale radius. Meanwhile, at low accretion rates, the BCG encroaches on this region and results in a total profile that is consistently steeper than $-2$. However, the disagreement we find between the stellar and dark matter $r_{\rm s}$ when fitting the orbiting profiles is not a result of this concealing of the peak at low $\Gamma$. Rather, it is likely a consequence of the suppressing of the satellite density around $r_{\rm s}$ by the angular median. This is discussed in more detail in Appendix~\ref{sec:appendix-a}.


\section{Discussion}
\label{sec:discussion}

\subsection{Comparison to Previous Simulations}
The agreement between the radius of steepest slope of the dark matter and stellar density in simulated galaxy clusters shown by \citet{Deason2021} suggests a close relationship between their splashback structure. We decomposed the stars and dark matter into their orbiting and infalling components and parameterized the edge of the orbiting profile by the truncation radius, $r_{\rm t}$, and slope, $\beta$ from \citet{Diemer2025}. We showed that the $r_{\rm t}$ of the stars and dark matter coincide, and that $\beta$ is generally larger for the stars. In other words, the splashback region is sharper for stars, but is at the same location as that of the dark matter.

This result is consistent with previous simulations. For example, \citet{Diemer2017} demonstrated that the splashback radius is a robust halo boundary across a wide range of masses, accretion rates, and cosmologies, and provided fitting functions that match our results well. \citet{ONeil.etal.2021} showed that baryonic physics has minimal impact on the location of the splashback radius, supporting our use of stellar profiles as reliable tracers. However, \citet{Lebeau.etal.2024} highlighted that the local cosmic web and associated accretion of galaxies from filaments can distort splashback measurements, especially in dynamically disturbed clusters like Virgo. These findings underscore the importance of accounting for environmental and dynamical diversity when interpreting splashback features.

The coincidence between the $r_{\rm t}$ means that the stellar $r_{\rm t}$--$\Gamma$ relation has the same form as that of the dark matter. This makes it a complimentary estimator of the accretion rate in clusters---assuming that $r_{\rm t}$ can be measured accurately. However, in observations, information of the orbiting and infalling components is inaccessible, and $r_{\rm t}$ must instead be estimated by fitting the total profile in projection. Given the freedom allowed by the full nine-parameter model of the total profile, the orbiting and infalling components may not always be accurately recovered.

When fitting the total profile, we find that the projected truncation radius, $R_{\rm t}$, has a standard deviation from the ``true'' $R_{\rm t}$ (measured from the orbiting profile) of $\sim 0.3\,R_{200{\rm m}}$ for both the dark matter and stars. There may also be a slight $R_{\rm t}$-dependent systematic difference in the case of the stars, which would produce a subtle change in the $R_{\rm t}$--$\Gamma$ relation, but this is unlikely to be discernable. While the exponential relationship can still be clearly seen, the scatter is larger than that of the $R_{\rm steep}$--$\Gamma$ relationship and so it is likely to be a poorer estimator of the accretion rate in observations.

\begin{figure*}
    \centering
    \includegraphics[width=0.8\linewidth]{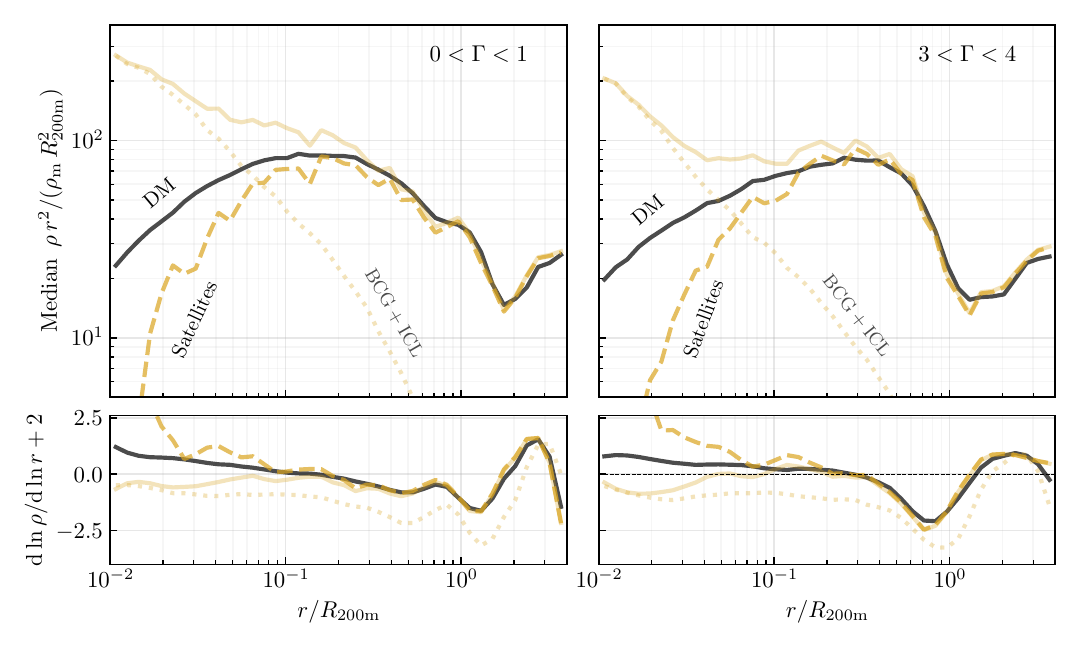}
    \caption{Median spherically-averaged $\rho r^2$ profiles of the dark matter and stars, in two bins of $\Gamma$. The stars have been separated into those contained in satellites---as determined by \textsc{ahf}---and all other stars, i.e. the smooth BCG+ICL. The stellar profiles have been rescaled $1.5\times M_{\rm DM}(<R_{200{\rm c}})/M_\ast(<R_{200{\rm c}})$ so that they approximately lie on top of the dark matter curves. The corresponding logarithmic slope profiles are shown in the bottom panel. We emphasise that the profiles that are being averaged here are the spherically-averaged profiles---they have not been binned in angular segments as is done for the other results in this work.}
    \label{fig:DM-stellar-profile-comparison-angmean}
\end{figure*}

\subsection{Comparison to Observations}\label{sec:comparison-to-observations}
Gravitational lensing and measurements of the distribution of cluster galaxies in projection provide complementary approaches to identifying the splashback boundary in galaxy clusters, each offering independent insight into the halo’s outer structure. Weak lensing directly probes the total matter distribution and has revealed a steepening in the density profile consistent with theoretical predictions of the splashback feature \citep[e.g.][]{Chang.etal.2018,Contigiani.etal.2019} that can be fit with an orbiting-infalling fitting function \citep[e.g.][]{Giocoli2024,Mpetha.etal.2025}. However, lensing measurements are inherently noisy at large radii due to shape noise and projection effects, and often require stacking to obtain robust profiles.

\citet{Chang.etal.2018} conducted one of the earliest joint analyses of galaxy density and weak lensing profiles using DES Year 1 data, detecting a splashback-like steepening in both tracers. They measured a splashback radius of \( r_{\mathrm{sp}} = 1.34 \pm 0.21\, h^{-1}\,\mathrm{Mpc} \), consistent with predictions from simulations and earlier SDSS-based studies. This dual detection reinforced the interpretation of splashback as a genuine feature of the matter distribution, rather than an artefact of galaxy selection or projection effects.
\citet{Mpetha.etal.2025} used weak lensing profiles from the UNIONS survey to constrain cosmological parameters via the infall region structure of galaxy clusters. Their analysis yielded splashback radii of \( 1.59^{+0.16}_{-0.13} \), \( 1.30^{+0.25}_{-0.13} \), and \( 1.45 \pm 0.11\, \mathrm{cMpc}/h \) for three independent cluster samples, and provided constraints of \( \Omega_{\mathrm{m}} = 0.29 \pm 0.05 \) and \( \sigma_8 = 0.80 \pm 0.04 \). These results demonstrate that splashback-based lensing analyses can break degeneracies inherent in traditional abundance and shear-based cosmological tests.

Lensing studies also highlight systematic challenges. \citet{Zhang.etal.2023} showed that neglecting the splashback feature in mass profile modelling can bias weak lensing mass estimates, particularly for group-sized halos (\( M_{200m} \sim 10^{13.5}\, M_\odot \)), by more than 0.1 dex. This bias propagates into cosmological analyses, potentially leading to underestimates of \( \Omega_{\mathrm{m}} \). \citet{Umetsu.Diemer.2017} further demonstrated that the splashback feature is most pronounced when cluster profiles are scaled by \( R_{200m} \), and that averaging over physical radii can smear out the steepening signal, underscoring the importance of profile scaling in lensing analyses.

Overall, lensing studies have established the splashback radius as a robust and observable feature of cluster mass profiles, with significant implications for halo boundary definitions and cosmological parameter estimation. As future surveys such as Euclid and LSST deliver deeper and wider lensing data, splashback-based analyses are poised to become a key tool in precision cosmology.

Cluster samples selected by their Sunyaev-Zel'dovich (SZ) effect offer a redshift-independent, mass-limited catalogue of clusters with minimal projection bias. Studies using clusters from Planck, SPT, and ACT have demonstrated that the splashback radius inferred from the galaxy distribution in SZ-selected samples closely matches expectations from simulations \citep[e.g.][]{Shin.etal.2019,Zurcher.More.2019,Adhikari.etal.2021}. In contrast, optically selected clusters—such as those in \textsc{redMaPPer} samples—often yield smaller splashback radii due to selection and projection effects \citep{Murata2020}. Because SZ selection is more tightly correlated with halo mass, it provides an important benchmark for testing the physical interpretation of the splashback feature.

Our analysis of the stellar splashback feature, traced through the total orbiting and infalling stellar density profiles, adds a new observational window on this boundary. We find that the truncation radius of the stellar orbiting component coincides with that of the dark matter and follows the same dependence on the mass accretion rate, consistent with earlier simulation-based studies \citep[][]{Deason2021,Diemer2025}. The steeper fall-off in the stellar profile makes the splashback edge more sharply defined than in lensing profiles, and may be observable through deep imaging of the ICL. 

Observations have not yet reached the surface-brightness limits required to trace the density profiles of galaxy clusters to the splashback radius, though rapid progress is being made. \citet{Deason2021} compared stellar profiles from the C-EAGLE simulations \citep{Barnes2017} to stacked surface brightness measurements of $z = 0.25$ DES Year-1 clusters \citep{Zhang2019}, demonstrating that depths of $\mu \simeq 32$--$36~{\rm mag\, arcsec^{-2}}$ are required to probe the splashback regime. These limits are expected to be reached in forthcoming surveys, including LSST \citep{Ivezic2019}.

Given the observational challenges, we have also considered measuring the splashback feature using galaxy number density profiles, which are more readily observable. Because the stellar mass in the outskirts of clusters is dominated by satellites, we expect---and find---very close agreement between the total stellar profile and galaxy number density profiles. We refer the interested reader to Appendix~\ref{sec:appendix-b}.

Our findings validate the use of projected stellar profiles as observational proxies for cluster boundaries. While subject to baryonic effects, the close alignment with dark matter splashback structures highlights the value of stellar tracers for probing cluster dynamics.

Taken together, lensing and stellar-based methods provide a powerful set of tools for characterising the outskirts of galaxy clusters. Our results, which show that the stellar splashback radius coincides with that of the dark matter and follows the same accretion rate dependence, suggest that stellar tracers—especially the ICL—can provide a complementary and potentially sharper observational probe of cluster boundaries, provided that projection effects and substructure contamination are properly accounted for.

Each is sensitive to different physical components—total mass, thermal pressure, and stellar material—and affected by distinct systematics. The convergence of splashback measurements across these methods strengthens the interpretation of this feature as a physically meaningful halo boundary and supports its use as a probe of cluster growth and accretion history \citep[e.g.][]{Joshi.etal.2026}.

\section{Conclusions}
\label{sec:conclusions}
We have used hydrodynamical simulations from The Three Hundred Project to study the splashback structure of galaxy clusters, focusing on the stellar and dark matter components. By decomposing each into orbiting and infalling material, we quantified their density profiles and characterised the splashback feature through the truncation radius $r_{\mathrm{t}}$ and slope $\beta$. Our key findings are as follows:

\begin{enumerate}
\item \emph{Coincidence of Splashback Radius:} The splashback radius $r_{\rm t}$ of the stellar component coincides with that of the dark matter on a halo-by-halo basis, confirming a close dynamical coupling between stars and dark matter in the cluster outskirts. This supports earlier results from \citet{Deason2021}, and extends them by demonstrating this agreement across a statistically representative sample of clusters.
\item\emph{Sharper Stellar Truncation:} The stellar orbiting profile exhibits a steeper truncation than the dark matter, consistent with a sharper splashback feature. This is in line with the findings of \citet{Deason2021} and \citet{ONeil.etal.2021}, and suggests that stellar material---in particular the ICL---may offer a more distinct observational signature of the splashback boundary.
\item\emph{Accretion Rate Dependence:} Both components follow the same $r_{\rm t}$--$\Gamma$ relation, enabling the use of stellar profiles as tracers of halo mass accretion history. This builds on the theoretical framework established by \citet{Adhikari2014}, \citet{More2015}, and \citet{Diemer2017}, and confirms that the splashback radius is sensitive to recent mass growth, with minimal dependence on halo mass or cosmology.
\item\emph{Observational Viability:} Projected stellar density profiles allow recovery of the splashback radius $R_{\rm t}$ with modest scatter ($\sim 0.3 R_{200m}$), demonstrating its observational viability. This complements observational detections of splashback features in galaxy density and weak lensing profiles (e.g. \citealt{Chang.etal.2018}; \citealt{Baxter.etal.2017}; \citealt{Gabriel-Silve.Sodre.2025}) and supports the use of the ICL as a tracer of cluster boundaries, especially in deep imaging surveys.
\item\emph{Infalling and Inner Profile Correlations:} The infalling stellar profiles correlate with those of the dark matter, and at high accretion rate the stellar profile reliably traces the dark matter near the scale radius $r_s$. This suggests that stellar material can be used to probe both the outer and inner dynamical structure of clusters, consistent with results from \citet{Pizzardo.etal.2024} and \citet{ONeil.etal.2021}.
\end{enumerate}

These results establish the stellar splashback feature—detectable via the ICL or integrated stellar profiles—as a robust, physically motivated tracer of cluster boundaries and growth rates. This opens a path for future studies to exploit stellar observables in constraining the dynamical state and assembly history of clusters, complementing traditional methods such as satellite kinematics and weak lensing. 

As upcoming surveys like Euclid, LSST, and Roman deliver deeper imaging and improved statistics, the stellar splashback radius may become a key tool in precision cluster cosmology. For this reason, there should be a focus on developing observational pipelines for ICL-based splashback detection, quantifying systematics due to projection and substructure, and integrating splashback measurements into multi-probe cosmological frameworks \citep[e.g.][]{Mpetha.etal.2025,Joshi.etal.2026}.

\section*{Acknowledgements}
KW and CP acknowledge the support of the ARC Centre of Excellence for All Sky Astrophysics in 3 Dimensions (ASTRO 3D), through project number CE170100013. AK is supported by the Spanish Ministerio de Ciencia e Innovación, (MICINN) under research grant PID2021-122603NB-C21 as well as project PID2024-156100NB-C21 financed by MICIU /AEI/10.13039/501100011033/FEDER, UE. AK further thanks Slowdive again, this time for 'everything is alive'. WC is supported by the STFC AGP Grant ST/V000594/1, the Atracci\'{o}n de Talento Contract no. 2020-T1/TIC-19882 was granted by the Comunidad de Madrid in Spain, and the science research grants were from the China Manned Space Project. He also thanks the Ministerio de Ciencia e Innovación (Spain) for financial support under Project grant PID2021-122603NB-C21 and HORIZON EUROPE Marie Sklodowska-Curie Actions for supporting the LACEGAL-III project with grant number 101086388.This work has been made possible by the {\small The Three Hundred} collaboration\footnote{\url{https://www.the300-project.org}}. The HD simulations (7K and 15K runs) were performed on the MareNostrum Finisterrae3, and Cibeles Supercomputers through The Red Española de Supercomputación grants (AECT-2022-3-0027, AECT-2023-1-0013, AECT-2023-2-0004, AECT-2023-3-0023, AECT-2024-1-0026), on the DIaL3 -- DiRAC Data Intensive service at the University of Leicester through the RAC15 grant: Seedcorn/ACTP317, and on the Niagara supercomputer at the SciNet HPC Consortium. DIaL3 is managed by the University of Leicester Research Computing Service on behalf of the STFC DiRAC HPC Facility (\url{https://www.dirac.ac.uk}). The DiRAC service at Leicester was funded by BEIS, UKRI and STFC capital funding and STFC operations grants. DiRAC is part of the UKRI Digital Research Infrastructure. This work also used the DiRAC Complexity system, operated by the University of Leicester IT Services, which forms part of the STFC DiRAC HPC Facility (\url{https://www.dirac.ac.uk}). This equipment is funded by BIS National E-Infrastructure capital grant ST/K000373/1 and STFC DiRAC Operations grant ST/K0003259/1. DiRAC is part of the National e-Infrastructure. SciNet \citep{Loken_2010} is funded by Innovation, Science and Economic Development Canada; the Digital Research Alliance of Canada; the Ontario Research Fund: Research Excellence; and the University of Toronto.

\section*{Data Availability}
All data used in our analysis are from {\small The Three Hundred} collaboration.

\bibliographystyle{mnras}
\bibliography{references} 

@ARTICLE{Diemer2014,
       author = {{Diemer}, Benedikt and {Kravtsov}, Andrey V.},
        title = "{Dependence of the Outer Density Profiles of Halos on Their Mass Accretion Rate}",
      journal = {\apj},
         year = 2014,
        month = jul,
       volume = {789},
       number = {1},
          eid = {1},
        pages = {1},
          doi = {10.1088/0004-637X/789/1/1},
archivePrefix = {arXiv},
       eprint = {1401.1216},
 primaryClass = {astro-ph.CO},
       adsurl = {https://ui.adsabs.harvard.edu/abs/2014ApJ...789....1D}
}

@ARTICLE{Adhikari2014,
       author = {{Adhikari}, Susmita and {Dalal}, Neal and {Chamberlain}, Robert T.},
        title = "{Splashback in accreting dark matter halos}",
      journal = {\jcap},
         year = 2014,
        month = nov,
       volume = {2014},
       number = {11},
        pages = {019-019},
          doi = {10.1088/1475-7516/2014/11/019},
archivePrefix = {arXiv},
       eprint = {1409.4482},
 primaryClass = {astro-ph.CO},
       adsurl = {https://ui.adsabs.harvard.edu/abs/2014JCAP...11..019A}
}

@ARTICLE{More2015,
       author = {{More}, Surhud and {Diemer}, Benedikt and {Kravtsov}, Andrey V.},
        title = "{The Splashback Radius as a Physical Halo Boundary and the Growth of Halo Mass}",
      journal = {\apj},
         year = 2015,
        month = sep,
       volume = {810},
       number = {1},
          eid = {36},
        pages = {36},
          doi = {10.1088/0004-637X/810/1/36},
archivePrefix = {arXiv},
       eprint = {1504.05591},
 primaryClass = {astro-ph.CO},
       adsurl = {https://ui.adsabs.harvard.edu/abs/2015ApJ...810...36M}
}

@ARTICLE{Diemer2017,
       author = {{Diemer}, Benedikt},
        title = "{The Splashback Radius of Halos from Particle Dynamics. I. The SPARTA Algorithm}",
      journal = {\apjs},
         year = 2017,
        month = jul,
       volume = {231},
       number = {1},
          eid = {5},
        pages = {5},
          doi = {10.3847/1538-4365/aa799c},
archivePrefix = {arXiv},
       eprint = {1703.09712},
 primaryClass = {astro-ph.CO},
       adsurl = {https://ui.adsabs.harvard.edu/abs/2017ApJS..231....5D}
}

@ARTICLE{More2016,
       author = {{More}, Surhud and {Miyatake}, Hironao and {Takada}, Masahiro and {Diemer}, Benedikt and {Kravtsov}, Andrey V. and {Dalal}, Neal K. and {More}, Anupreeta and {Murata}, Ryoma and {Mandelbaum}, Rachel and {Rozo}, Eduardo and {Rykoff}, Eli S. and {Oguri}, Masamune and {Spergel}, David N.},
        title = "{Detection of the Splashback Radius and Halo Assembly Bias of Massive Galaxy Clusters}",
      journal = {\apj},
         year = 2016,
        month = jul,
       volume = {825},
       number = {1},
          eid = {39},
        pages = {39},
          doi = {10.3847/0004-637X/825/1/39},
archivePrefix = {arXiv},
       eprint = {1601.06063},
 primaryClass = {astro-ph.CO},
       adsurl = {https://ui.adsabs.harvard.edu/abs/2016ApJ...825...39M}
}

@ARTICLE{Chang2018,
       author = {{Chang}, C. and {Baxter}, E. and {Jain}, B. and {S{\'a}nchez}, C. and {Adhikari}, S. and {Varga}, T.~N. and {Fang}, Y. and {Rozo}, E. and {Rykoff}, E.~S. and {Kravtsov}, A. and {Gruen}, D. and {Hartley}, W. and {Huff}, E.~M. and {Jarvis}, M. and {Kim}, A.~G. and {Prat}, J. and {MacCrann}, N. and {McClintock}, T. and {Palmese}, A. and {Rapetti}, D. and {Rollins}, R.~P. and {Samuroff}, S. and {Sheldon}, E. and {Troxel}, M.~A. and {Wechsler}, R.~H. and {Zhang}, Y. and {Zuntz}, J. and {Abbott}, T.~M.~C. and {Abdalla}, F.~B. and {Allam}, S. and {Annis}, J. and {Bechtol}, K. and {Benoit-L{\'e}vy}, A. and {Bernstein}, G.~M. and {Brooks}, D. and {Buckley-Geer}, E. and {Carnero Rosell}, A. and {Carrasco Kind}, M. and {Carretero}, J. and {D'Andrea}, C.~B. and {da Costa}, L.~N. and {Davis}, C. and {Desai}, S. and {Diehl}, H.~T. and {Dietrich}, J.~P. and {Drlica-Wagner}, A. and {Eifler}, T.~F. and {Flaugher}, B. and {Fosalba}, P. and {Frieman}, J. and {Garc{\'\i}a-Bellido}, J. and {Gaztanaga}, E. and {Gerdes}, D.~W. and {Gruendl}, R.~A. and {Gschwend}, J. and {Gutierrez}, G. and {Honscheid}, K. and {James}, D.~J. and {Jeltema}, T. and {Krause}, E. and {Kuehn}, K. and {Lahav}, O. and {Lima}, M. and {March}, M. and {Marshall}, J.~L. and {Martini}, P. and {Melchior}, P. and {Menanteau}, F. and {Miquel}, R. and {Mohr}, J.~J. and {Nord}, B. and {Ogando}, R.~L.~C. and {Plazas}, A.~A. and {Sanchez}, E. and {Scarpine}, V. and {Schindler}, R. and {Schubnell}, M. and {Sevilla-Noarbe}, I. and {Smith}, M. and {Smith}, R.~C. and {Soares-Santos}, M. and {Sobreira}, F. and {Suchyta}, E. and {Swanson}, M.~E.~C. and {Tarle}, G. and {Weller}, J. and {DES Collaboration}},
        title = "{The Splashback Feature around DES Galaxy Clusters: Galaxy Density and Weak Lensing Profiles}",
      journal = {\apj},
         year = 2018,
        month = sep,
       volume = {864},
       number = {1},
          eid = {83},
        pages = {83},
          doi = {10.3847/1538-4357/aad5e7},
archivePrefix = {arXiv},
       eprint = {1710.06808},
 primaryClass = {astro-ph.CO},
       adsurl = {https://ui.adsabs.harvard.edu/abs/2018ApJ...864...83C}
}

@ARTICLE{Diemer2022,
       author = {{Diemer}, Benedikt},
        title = "{A dynamics-based density profile for dark haloes - I. Algorithm and basic results}",
      journal = {\mnras},
         year = 2022,
        month = jun,
       volume = {513},
       number = {1},
        pages = {573-594},
          doi = {10.1093/mnras/stac878},
archivePrefix = {arXiv},
       eprint = {2112.03921},
 primaryClass = {astro-ph.CO},
       adsurl = {https://ui.adsabs.harvard.edu/abs/2022MNRAS.513..573D}
}

@ARTICLE{Planck2016,
       author = {{Planck Collaboration} and {Ade}, P.~A.~R. and {Aghanim}, N. and {Arnaud}, M. and {Ashdown}, M. and {Aumont}, J. and {Baccigalupi}, C. and {Banday}, A.~J. and {Barreiro}, R.~B. and {Bartlett}, J.~G. and {Bartolo}, N. and {Battaner}, E. and {Battye}, R. and {Benabed}, K. and {Beno{\^\i}t}, A. and {Benoit-L{\'e}vy}, A. and {Bernard}, J. -P. and {Bersanelli}, M. and {Bielewicz}, P. and {Bock}, J.~J. and {Bonaldi}, A. and {Bonavera}, L. and {Bond}, J.~R. and {Borrill}, J. and {Bouchet}, F.~R. and {Boulanger}, F. and {Bucher}, M. and {Burigana}, C. and {Butler}, R.~C. and {Calabrese}, E. and {Cardoso}, J. -F. and {Catalano}, A. and {Challinor}, A. and {Chamballu}, A. and {Chary}, R. -R. and {Chiang}, H.~C. and {Chluba}, J. and {Christensen}, P.~R. and {Church}, S. and {Clements}, D.~L. and {Colombi}, S. and {Colombo}, L.~P.~L. and {Combet}, C. and {Coulais}, A. and {Crill}, B.~P. and {Curto}, A. and {Cuttaia}, F. and {Danese}, L. and {Davies}, R.~D. and {Davis}, R.~J. and {de Bernardis}, P. and {de Rosa}, A. and {de Zotti}, G. and {Delabrouille}, J. and {D{\'e}sert}, F. -X. and {Di Valentino}, E. and {Dickinson}, C. and {Diego}, J.~M. and {Dolag}, K. and {Dole}, H. and {Donzelli}, S. and {Dor{\'e}}, O. and {Douspis}, M. and {Ducout}, A. and {Dunkley}, J. and {Dupac}, X. and {Efstathiou}, G. and {Elsner}, F. and {En{\ss}lin}, T.~A. and {Eriksen}, H.~K. and {Farhang}, M. and {Fergusson}, J. and {Finelli}, F. and {Forni}, O. and {Frailis}, M. and {Fraisse}, A.~A. and {Franceschi}, E. and {Frejsel}, A. and {Galeotta}, S. and {Galli}, S. and {Ganga}, K. and {Gauthier}, C. and {Gerbino}, M. and {Ghosh}, T. and {Giard}, M. and {Giraud-H{\'e}raud}, Y. and {Giusarma}, E. and {Gjerl{\o}w}, E. and {Gonz{\'a}lez-Nuevo}, J. and {G{\'o}rski}, K.~M. and {Gratton}, S. and {Gregorio}, A. and {Gruppuso}, A. and {Gudmundsson}, J.~E. and {Hamann}, J. and {Hansen}, F.~K. and {Hanson}, D. and {Harrison}, D.~L. and {Helou}, G. and {Henrot-Versill{\'e}}, S. and {Hern{\'a}ndez-Monteagudo}, C. and {Herranz}, D. and {Hildebrandt}, S.~R. and {Hivon}, E. and {Hobson}, M. and {Holmes}, W.~A. and {Hornstrup}, A. and {Hovest}, W. and {Huang}, Z. and {Huffenberger}, K.~M. and {Hurier}, G. and {Jaffe}, A.~H. and {Jaffe}, T.~R. and {Jones}, W.~C. and {Juvela}, M. and {Keih{\"a}nen}, E. and {Keskitalo}, R. and {Kisner}, T.~S. and {Kneissl}, R. and {Knoche}, J. and {Knox}, L. and {Kunz}, M. and {Kurki-Suonio}, H. and {Lagache}, G. and {L{\"a}hteenm{\"a}ki}, A. and {Lamarre}, J. -M. and {Lasenby}, A. and {Lattanzi}, M. and {Lawrence}, C.~R. and {Leahy}, J.~P. and {Leonardi}, R. and {Lesgourgues}, J. and {Levrier}, F. and {Lewis}, A. and {Liguori}, M. and {Lilje}, P.~B. and {Linden-V{\o}rnle}, M. and {L{\'o}pez-Caniego}, M. and {Lubin}, P.~M. and {Mac{\'\i}as-P{\'e}rez}, J.~F. and {Maggio}, G. and {Maino}, D. and {Mandolesi}, N. and {Mangilli}, A. and {Marchini}, A. and {Maris}, M. and {Martin}, P.~G. and {Martinelli}, M. and {Mart{\'\i}nez-Gonz{\'a}lez}, E. and {Masi}, S. and {Matarrese}, S. and {McGehee}, P. and {Meinhold}, P.~R. and {Melchiorri}, A. and {Melin}, J. -B. and {Mendes}, L. and {Mennella}, A. and {Migliaccio}, M. and {Millea}, M. and {Mitra}, S. and {Miville-Desch{\^e}nes}, M. -A. and {Moneti}, A. and {Montier}, L. and {Morgante}, G. and {Mortlock}, D. and {Moss}, A. and {Munshi}, D. and {Murphy}, J.~A. and {Naselsky}, P. and {Nati}, F. and {Natoli}, P. and {Netterfield}, C.~B. and {N{\o}rgaard-Nielsen}, H.~U. and {Noviello}, F. and {Novikov}, D. and {Novikov}, I. and {Oxborrow}, C.~A. and {Paci}, F. and {Pagano}, L. and {Pajot}, F. and {Paladini}, R. and {Paoletti}, D. and {Partridge}, B. and {Pasian}, F. and {Patanchon}, G. and {Pearson}, T.~J. and {Perdereau}, O. and {Perotto}, L. and {Perrotta}, F. and {Pettorino}, V. and {Piacentini}, F. and {Piat}, M. and {Pierpaoli}, E. and {Pietrobon}, D. and {Plaszczynski}, S. and {Pointecouteau}, E. and {Polenta}, G. and {Popa}, L. and {Pratt}, G.~W. and {Pr{\'e}zeau}, G.},
        title = "{Planck 2015 results. XIII. Cosmological parameters}",
      journal = {\aap},
         year = 2016,
        month = sep,
       volume = {594},
          eid = {A13},
        pages = {A13},
          doi = {10.1051/0004-6361/201525830},
archivePrefix = {arXiv},
       eprint = {1502.01589},
 primaryClass = {astro-ph.CO},
       adsurl = {https://ui.adsabs.harvard.edu/abs/2016A&A...594A..13P}
}

@ARTICLE{Diemer2023,
       author = {{Diemer}, Benedikt},
        title = "{A dynamics-based density profile for dark haloes - II. Fitting function}",
      journal = {\mnras},
         year = 2023,
        month = mar,
       volume = {519},
       number = {3},
        pages = {3292-3311},
          doi = {10.1093/mnras/stac3778},
archivePrefix = {arXiv},
       eprint = {2205.03420},
 primaryClass = {astro-ph.CO},
       adsurl = {https://ui.adsabs.harvard.edu/abs/2023MNRAS.519.3292D}
}

@ARTICLE{Diemer2025,
       author = {{Diemer}, Benedikt},
        title = "{A dynamics-based density profile for dark haloes - III. Parameter space}",
      journal = {\mnras},
         year = 2025,
        month = jan,
       volume = {536},
       number = {2},
        pages = {1718-1735},
          doi = {10.1093/mnras/stae2717},
archivePrefix = {arXiv},
       eprint = {2410.17324},
 primaryClass = {astro-ph.CO},
       adsurl = {https://ui.adsabs.harvard.edu/abs/2025MNRAS.536.1718D}
}

@ARTICLE{Deason2021,
       author = {{Deason}, Alis J. and {Oman}, Kyle A. and {Fattahi}, Azadeh and {Schaller}, Matthieu and {Jauzac}, Mathilde and {Zhang}, Yuanyuan and {Montes}, Mireia and {Bah{\'e}}, Yannick M. and {Dalla Vecchia}, Claudio and {Kay}, Scott T. and {Evans}, Tilly A.},
        title = "{Stellar splashback: the edge of the intracluster light}",
      journal = {\mnras},
         year = 2021,
        month = jan,
       volume = {500},
       number = {3},
        pages = {4181-4192},
          doi = {10.1093/mnras/staa3590},
archivePrefix = {arXiv},
       eprint = {2010.02937},
 primaryClass = {astro-ph.GA},
       adsurl = {https://ui.adsabs.harvard.edu/abs/2021MNRAS.500.4181D}
}

@ARTICLE{Knollmann2009,
       author = {{Knollmann}, Steffen R. and {Knebe}, Alexander},
        title = "{AHF: Amiga's Halo Finder}",
      journal = {\apjs},
         year = 2009,
        month = jun,
       volume = {182},
       number = {2},
        pages = {608-624},
          doi = {10.1088/0067-0049/182/2/608},
archivePrefix = {arXiv},
       eprint = {0904.3662},
 primaryClass = {astro-ph.CO},
       adsurl = {https://ui.adsabs.harvard.edu/abs/2009ApJS..182..608K}
}

@ARTICLE{Joshi.etal.2026,
       author = {{Joshi}, Jitendra and {Rana}, Divya and {More}, Surhud and {Klein}, Matthias},
        title = "{Splashback radius and the mass accretion rate of RASS MCMF galaxy clusters}",
      journal = {\prd},
         year = 2026,
        month = jan,
       volume = {113},
       number = {2},
          eid = {023512},
        pages = {023512},
          doi = {10.1103/7d62-3gnp},
archivePrefix = {arXiv},
       eprint = {2506.08925},
 primaryClass = {astro-ph.CO},
       adsurl = {https://ui.adsabs.harvard.edu/abs/2026PhRvD.113b3512J}
}

@ARTICLE{Adhikari.etal.2021,
       author = {{Adhikari}, Susmita and {Shin}, Tae-hyeon and {Jain}, Bhuvnesh and {Hilton}, Matt and {Baxter}, Eric and {Chang}, Chihway and {Wechsler}, Risa H. and {Battaglia}, Nick and {Bond}, J. Richard and {Bocquet}, Sebastian and {Choi}, Steve K. and {DeRose}, Joseph and {Devlin}, Mark and {Dunkley}, Jo and {Evrard}, August E. and {Ferraro}, Simone and {Hill}, J. Colin and {Hughes}, John P. and {Gallardo}, Patricio A. and {Lokken}, Martine and {MacInnis}, Amanda and {Madhavacheril}, Mathew S. and {McMahon}, Jeffrey and {Nati}, Frederico and {Newburgh}, Laura B. and {Niemack}, Michael D. and {Page}, Lyman A. and {Palmese}, Antonella and {Partridge}, Bruce and {Rozo}, Eduardo and {Rykoff}, Eli and {Salatino}, Maria and {Schillaci}, Alessandro and {Sehgal}, Neelima and {Sif{\'o}n}, Crist{\'o}bal and {To}, Chun-Hao and {Wollack}, Ed and {Wu}, Hao-Yi and {Xu}, Zhilei and {Aguena}, Michel and {Allam}, Sahar and {Amon}, Alexandra and {Annis}, James and {Avila}, Santiago and {Bacon}, David and {Bertin}, Emmanuel and {Bhargava}, Sunayana and {Brooks}, David and {Burke}, David L. and {Rosell}, Aurelio C. and {Kind}, Matias Carrasco and {Carretero}, Jorge and {Castander}, Francisco Javier and {Choi}, Ami and {Costanzi}, Matteo and {da Costa}, Luiz N. and {Vicente}, Juan De and {Desai}, Shantanu and {Diehl}, Thomas H. and {Doel}, Peter and {Everett}, Spencer and {Ferrero}, Ismael and {Fert{\'e}}, Agn{\`e}s and {Flaugher}, Brenna and {Fosalba}, Pablo and {Frieman}, Josh and {Garc{\'\i}a-Bellido}, Juan and {Gaztanaga}, Enrique and {Gruen}, Daniel and {Gruendl}, Robert A. and {Gschwend}, Julia and {Gutierrez}, Gaston and {Hartley}, Will G. and {Hinton}, Samuel R. and {Hollowood}, Devon L. and {Honscheid}, Klaus and {James}, David J. and {Jeltema}, Tesla and {Kuehn}, Kyler and {Kuropatkin}, Nikolay and {Lahav}, Ofer and {Lima}, Marcos and {Maia}, Marcio A.~G. and {Marshall}, Jennifer L. and {Martini}, Paul and {Melchior}, Peter and {Menanteau}, Felipe and {Miquel}, Ramon and {Morgan}, Robert and {L.~C. Ogando}, Ricardo and {Paz-Chinch{\'o}n}, Francisco and {Malag{\'o}n}, Andr{\'e}s Plazas and {Sanchez}, Eusebio and {Santiago}, Basilio and {Scarpine}, Vic and {Serrano}, Santiago and {Sevilla-Noarbe}, Ignacio and {Smith}, Mathew and {Soares-Santos}, Marcelle and {Suchyta}, Eric and {E.~C. Swanson}, Molly and {Varga}, Tamas N. and {Wilkinson}, Reese D. and {Zhang}, Yuanyuan and {Austermann}, Jason E. and {Beall}, James A. and {Becker}, Daniel T. and {Denison}, Edward V. and {Duff}, Shannon M. and {Hilton}, Gene C. and {Hubmayr}, Johannes and {Ullom}, Joel N. and {Lanen}, Jeff Van and {Vale}, Leila R. and {Vale}, Leila R. and {Vale}, Leila R.},
        title = "{Probing Galaxy Evolution in Massive Clusters Using ACT and DES: Splashback as a Cosmic Clock}",
      journal = {\apj},
         year = 2021,
        month = dec,
       volume = {923},
       number = {1},
          eid = {37},
        pages = {37},
          doi = {10.3847/1538-4357/ac0bbc},
archivePrefix = {arXiv},
       eprint = {2008.11663},
 primaryClass = {astro-ph.GA},
       adsurl = {https://ui.adsabs.harvard.edu/abs/2021ApJ...923...37A}
}

@ARTICLE{Shin.etal.2019,
       author = {{Shin}, T. and {Adhikari}, S. and {Baxter}, E.~J. and {Chang}, C. and {Jain}, B. and {Battaglia}, N. and {Bleem}, L. and {Bocquet}, S. and {DeRose}, J. and {Gruen}, D. and {Hilton}, M. and {Kravtsov}, A. and {McClintock}, T. and {Rozo}, E. and {Rykoff}, E.~S. and {Varga}, T.~N. and {Wechsler}, R.~H. and {Wu}, H. and {Zhang}, Z. and {Aiola}, S. and {Allam}, S. and {Bechtol}, K. and {Benson}, B.~A. and {Bertin}, E. and {Bond}, J.~R. and {Brodwin}, M. and {Brooks}, D. and {Buckley-Geer}, E. and {Burke}, D.~L. and {Carlstrom}, J.~E. and {Carnero Rosell}, A. and {Carrasco Kind}, M. and {Carretero}, J. and {Castander}, F.~J. and {Choi}, S.~K. and {Cunha}, C.~E. and {Crawford}, T.~M. and {da Costa}, L.~N. and {De Vicente}, J. and {Desai}, S. and {Devlin}, M.~J. and {Dietrich}, J.~P. and {Doel}, P. and {Dunkley}, J. and {Eifler}, T.~F. and {Evrard}, A.~E. and {Flaugher}, B. and {Fosalba}, P. and {Gallardo}, P.~A. and {Garc{\'\i}a-Bellido}, J. and {Gaztanaga}, E. and {Gerdes}, D.~W. and {Gralla}, M. and {Gruendl}, R.~A. and {Gschwend}, J. and {Gupta}, N. and {Gutierrez}, G. and {Hartley}, W.~G. and {Hill}, J.~C. and {Ho}, S.~P. and {Hollowood}, D.~L. and {Honscheid}, K. and {Hoyle}, B. and {Huffenberger}, K. and {Hughes}, J.~P. and {James}, D.~J. and {Jeltema}, T. and {Kim}, A.~G. and {Krause}, E. and {Kuehn}, K. and {Lahav}, O. and {Lima}, M. and {Madhavacheril}, M.~S. and {Maia}, M.~A.~G. and {Marshall}, J.~L. and {Maurin}, L. and {McMahon}, J. and {Menanteau}, F. and {Miller}, C.~J. and {Miquel}, R. and {Mohr}, J.~J. and {Naess}, S. and {Nati}, F. and {Newburgh}, L. and {Niemack}, M.~D. and {Ogando}, R.~L.~C. and {Page}, L.~A. and {Partridge}, B. and {Patil}, S. and {Plazas}, A.~A. and {Rapetti}, D. and {Reichardt}, C.~L. and {Romer}, A.~K. and {Sanchez}, E. and {Scarpine}, V. and {Schindler}, R. and {Serrano}, S. and {Smith}, M. and {Smith}, R.~C. and {Soares-Santos}, M. and {Sobreira}, F. and {Staggs}, S.~T. and {Stark}, A. and {Stein}, G. and {Suchyta}, E. and {Swanson}, M.~E.~C. and {Tarle}, G. and {Thomas}, D. and {van Engelen}, A. and {Wollack}, E.~J. and {Xu}, Z.},
        title = "{Measurement of the splashback feature around SZ-selected Galaxy clusters with DES, SPT, and ACT}",
      journal = {\mnras},
         year = 2019,
        month = aug,
       volume = {487},
       number = {2},
        pages = {2900-2918},
          doi = {10.1093/mnras/stz1434},
archivePrefix = {arXiv},
       eprint = {1811.06081},
 primaryClass = {astro-ph.CO},
       adsurl = {https://ui.adsabs.harvard.edu/abs/2019MNRAS.487.2900S}
}

@ARTICLE{Contigiani.etal.2019,
       author = {{Contigiani}, Omar and {Hoekstra}, Henk and {Bah{\'e}}, Yannick M.},
        title = "{Weak lensing constraints on splashback around massive clusters}",
      journal = {\mnras},
         year = 2019,
        month = may,
       volume = {485},
       number = {1},
        pages = {408-415},
          doi = {10.1093/mnras/stz404},
archivePrefix = {arXiv},
       eprint = {1809.10045},
 primaryClass = {astro-ph.CO},
       adsurl = {https://ui.adsabs.harvard.edu/abs/2019MNRAS.485..408C}
}

@ARTICLE{Zurcher.More.2019,
       author = {{Z{\"u}rcher}, Dominik and {More}, Surhud},
        title = "{The Splashback Radius of Planck SZ Clusters}",
      journal = {\apj},
         year = 2019,
        month = apr,
       volume = {874},
       number = {2},
          eid = {184},
        pages = {184},
          doi = {10.3847/1538-4357/ab08e8},
archivePrefix = {arXiv},
       eprint = {1811.06511},
 primaryClass = {astro-ph.CO},
       adsurl = {https://ui.adsabs.harvard.edu/abs/2019ApJ...874..184Z}
}

@ARTICLE{Chang.etal.2018,
       author = {{Chang}, C. and {Baxter}, E. and {Jain}, B. and {S{\'a}nchez}, C. and {Adhikari}, S. and {Varga}, T.~N. and {Fang}, Y. and {Rozo}, E. and {Rykoff}, E.~S. and {Kravtsov}, A. and {Gruen}, D. and {Hartley}, W. and {Huff}, E.~M. and {Jarvis}, M. and {Kim}, A.~G. and {Prat}, J. and {MacCrann}, N. and {McClintock}, T. and {Palmese}, A. and {Rapetti}, D. and {Rollins}, R.~P. and {Samuroff}, S. and {Sheldon}, E. and {Troxel}, M.~A. and {Wechsler}, R.~H. and {Zhang}, Y. and {Zuntz}, J. and {Abbott}, T.~M.~C. and {Abdalla}, F.~B. and {Allam}, S. and {Annis}, J. and {Bechtol}, K. and {Benoit-L{\'e}vy}, A. and {Bernstein}, G.~M. and {Brooks}, D. and {Buckley-Geer}, E. and {Carnero Rosell}, A. and {Carrasco Kind}, M. and {Carretero}, J. and {D'Andrea}, C.~B. and {da Costa}, L.~N. and {Davis}, C. and {Desai}, S. and {Diehl}, H.~T. and {Dietrich}, J.~P. and {Drlica-Wagner}, A. and {Eifler}, T.~F. and {Flaugher}, B. and {Fosalba}, P. and {Frieman}, J. and {Garc{\'\i}a-Bellido}, J. and {Gaztanaga}, E. and {Gerdes}, D.~W. and {Gruendl}, R.~A. and {Gschwend}, J. and {Gutierrez}, G. and {Honscheid}, K. and {James}, D.~J. and {Jeltema}, T. and {Krause}, E. and {Kuehn}, K. and {Lahav}, O. and {Lima}, M. and {March}, M. and {Marshall}, J.~L. and {Martini}, P. and {Melchior}, P. and {Menanteau}, F. and {Miquel}, R. and {Mohr}, J.~J. and {Nord}, B. and {Ogando}, R.~L.~C. and {Plazas}, A.~A. and {Sanchez}, E. and {Scarpine}, V. and {Schindler}, R. and {Schubnell}, M. and {Sevilla-Noarbe}, I. and {Smith}, M. and {Smith}, R.~C. and {Soares-Santos}, M. and {Sobreira}, F. and {Suchyta}, E. and {Swanson}, M.~E.~C. and {Tarle}, G. and {Weller}, J. and {DES Collaboration}},
        title = "{The Splashback Feature around DES Galaxy Clusters: Galaxy Density and Weak Lensing Profiles}",
      journal = {\apj},
         year = 2018,
        month = sep,
       volume = {864},
       number = {1},
          eid = {83},
        pages = {83},
          doi = {10.3847/1538-4357/aad5e7},
archivePrefix = {arXiv},
       eprint = {1710.06808},
 primaryClass = {astro-ph.CO},
       adsurl = {https://ui.adsabs.harvard.edu/abs/2018ApJ...864...83C}
}

@ARTICLE{Cui.etal.2018,
       author = {{Cui}, Weiguang and {Knebe}, Alexander and {Yepes}, Gustavo and {Pearce}, Frazer and {Power}, Chris and {Dave}, Romeel and {Arth}, Alexander and {Borgani}, Stefano and {Dolag}, Klaus and {Elahi}, Pascal and {Mostoghiu}, Robert and {Murante}, Giuseppe and {Rasia}, Elena and {Stoppacher}, Doris and {Vega-Ferrero}, Jesus and {Wang}, Yang and {Yang}, Xiaohu and {Benson}, Andrew and {Cora}, Sof{\'\i}a A. and {Croton}, Darren J. and {Sinha}, Manodeep and {Stevens}, Adam R.~H. and {Vega-Mart{\'\i}nez}, Cristian A. and {Arthur}, Jake and {Baldi}, Anna S. and {Ca{\~n}as}, Rodrigo and {Cialone}, Giammarco and {Cunnama}, Daniel and {De Petris}, Marco and {Durando}, Giacomo and {Ettori}, Stefano and {Gottl{\"o}ber}, Stefan and {Nuza}, Sebasti{\'a}n E. and {Old}, Lyndsay J. and {Pilipenko}, Sergey and {Sorce}, Jenny G. and {Welker}, Charlotte},
        title = "{The Three Hundred project: a large catalogue of theoretically modelled galaxy clusters for cosmological and astrophysical applications}",
      journal = {\mnras},
         year = 2018,
        month = nov,
       volume = {480},
       number = {3},
        pages = {2898-2915},
          doi = {10.1093/mnras/sty2111},
archivePrefix = {arXiv},
       eprint = {1809.04622},
 primaryClass = {astro-ph.GA},
       adsurl = {https://ui.adsabs.harvard.edu/abs/2018MNRAS.480.2898C}
}

@ARTICLE{Cui.etal.2022,
       author = {{Cui}, Weiguang and {Dave}, Romeel and {Knebe}, Alexander and {Rasia}, Elena and {Gray}, Meghan and {Pearce}, Frazer and {Power}, Chris and {Yepes}, Gustavo and {Anbajagane}, Dhayaa and {Ceverino}, Daniel and {Contreras-Santos}, Ana and {de Andres}, Daniel and {De Petris}, Marco and {Ettori}, Stefano and {Haggar}, Roan and {Li}, Qingyang and {Wang}, Yang and {Yang}, Xiaohu and {Borgani}, Stefano and {Dolag}, Klaus and {Zu}, Ying and {Kuchner}, Ulrike and {Ca{\~n}as}, Rodrigo and {Ferragamo}, Antonio and {Gianfagna}, Giulia},
        title = "{THE THREE HUNDRED project: The GIZMO-SIMBA run}",
      journal = {\mnras},
         year = 2022,
        month = jul,
       volume = {514},
       number = {1},
        pages = {977-996},
          doi = {10.1093/mnras/stac1402},
archivePrefix = {arXiv},
       eprint = {2202.14038},
 primaryClass = {astro-ph.GA},
       adsurl = {https://ui.adsabs.harvard.edu/abs/2022MNRAS.514..977C}
}

@ARTICLE{Dave.etal.2019,
       author = {{Dav{\'e}}, Romeel and {Angl{\'e}s-Alc{\'a}zar}, Daniel and {Narayanan}, Desika and {Li}, Qi and {Rafieferantsoa}, Mika H. and {Appleby}, Sarah},
        title = "{SIMBA: Cosmological simulations with black hole growth and feedback}",
      journal = {\mnras},
         year = 2019,
        month = jun,
       volume = {486},
       number = {2},
        pages = {2827-2849},
          doi = {10.1093/mnras/stz937},
archivePrefix = {arXiv},
       eprint = {1901.10203},
 primaryClass = {astro-ph.GA},
       adsurl = {https://ui.adsabs.harvard.edu/abs/2019MNRAS.486.2827D}
}

@ARTICLE{Klypin.etal.2016,
       author = {{Klypin}, Anatoly and {Yepes}, Gustavo and {Gottl{\"o}ber}, Stefan and {Prada}, Francisco and {He{\ss}}, Steffen},
        title = "{MultiDark simulations: the story of dark matter halo concentrations and density profiles}",
      journal = {\mnras},
         year = 2016,
        month = apr,
       volume = {457},
       number = {4},
        pages = {4340-4359},
          doi = {10.1093/mnras/stw248},
archivePrefix = {arXiv},
       eprint = {1411.4001},
 primaryClass = {astro-ph.CO},
       adsurl = {https://ui.adsabs.harvard.edu/abs/2016MNRAS.457.4340K}
}

@ARTICLE{Hopkins.2015,
       author = {{Hopkins}, Philip F.},
        title = "{A new class of accurate, mesh-free hydrodynamic simulation methods}",
      journal = {\mnras},
         year = 2015,
        month = jun,
       volume = {450},
       number = {1},
        pages = {53-110},
          doi = {10.1093/mnras/stv195},
archivePrefix = {arXiv},
       eprint = {1409.7395},
 primaryClass = {astro-ph.CO},
       adsurl = {https://ui.adsabs.harvard.edu/abs/2015MNRAS.450...53H}
}

@ARTICLE{Springel.2005,
       author = {{Springel}, Volker},
        title = "{The cosmological simulation code GADGET-2}",
      journal = {\mnras},
         year = 2005,
        month = dec,
       volume = {364},
       number = {4},
        pages = {1105-1134},
          doi = {10.1111/j.1365-2966.2005.09655.x},
archivePrefix = {arXiv},
       eprint = {astro-ph/0505010},
 primaryClass = {astro-ph},
       adsurl = {https://ui.adsabs.harvard.edu/abs/2005MNRAS.364.1105S}
}

@ARTICLE{Baxter.etal.2017,
       author = {{Baxter}, Eric and {Chang}, Chihway and {Jain}, Bhuvnesh and {Adhikari}, Susmita and {Dalal}, Neal and {Kravtsov}, Andrey and {More}, Surhud and {Rozo}, Eduardo and {Rykoff}, Eli and {Sheth}, Ravi K.},
        title = "{The Halo Boundary of Galaxy Clusters in the SDSS}",
      journal = {\apj},
         year = 2017,
        month = may,
       volume = {841},
       number = {1},
          eid = {18},
        pages = {18},
          doi = {10.3847/1538-4357/aa6ff0},
archivePrefix = {arXiv},
       eprint = {1702.01722},
 primaryClass = {astro-ph.CO},
       adsurl = {https://ui.adsabs.harvard.edu/abs/2017ApJ...841...18B}
}

@ARTICLE{Umetsu.Diemer.2017,
       author = {{Umetsu}, Keiichi and {Diemer}, Benedikt},
        title = "{Lensing Constraints on the Mass Profile Shape and the Splashback Radius of Galaxy Clusters}",
      journal = {\apj},
         year = 2017,
        month = feb,
       volume = {836},
       number = {2},
          eid = {231},
        pages = {231},
          doi = {10.3847/1538-4357/aa5c90},
archivePrefix = {arXiv},
       eprint = {1611.09366},
 primaryClass = {astro-ph.CO},
       adsurl = {https://ui.adsabs.harvard.edu/abs/2017ApJ...836..231U}
}

@ARTICLE{Mpetha.etal.2025,
       author = {{Mpetha}, C.~T. and {Taylor}, J.~E. and {Amoura}, Y. and {Haggar}, R. and {de Boer}, T. and {Guerrini}, S. and {Guinot}, A. and {Peters}, F. Hervas and {Hildebrandt}, H. and {Hudson}, M.~J. and {Kilbinger}, M. and {Liaudat}, T. and {McConnachie}, A. and {Van Waerbeke}, L. and {Wittje}, A.},
        title = "{Cosmology from UNIONS weak lensing profiles of galaxy clusters}",
      journal = {\mnras},
         year = 2025,
        month = oct,
       volume = {543},
       number = {2},
        pages = {1393-1409},
          doi = {10.1093/mnras/staf1538},
archivePrefix = {arXiv},
       eprint = {2501.09147},
 primaryClass = {astro-ph.CO},
       adsurl = {https://ui.adsabs.harvard.edu/abs/2025MNRAS.543.1393M}
}

@INPROCEEDINGS{Zhang.etal.2023,
       author = {{Zhang}, Congyao and {Zhuravleva}, Irina and {Churazov}, Eugene and {Kravtsov}, Andrey and {Dolag}, Klaus and {Forman}, William},
        title = "{Evolution of Shocks and Splashback Boundaries in Cluster Outskirts}",
    booktitle = {AAS/High Energy Astrophysics Division},
         year = 2023,
       series = {AAS/High Energy Astrophysics Division},
       volume = {20},
        month = sep,
          eid = {101.05},
        pages = {101.05},
       adsurl = {https://ui.adsabs.harvard.edu/abs/2023HEAD...2010105Z}
}

@ARTICLE{Ludlow2019,
       author = {{Ludlow}, Aaron D. and {Schaye}, Joop and {Bower}, Richard},
        title = "{Numerical convergence of simulations of galaxy formation: the abundance and internal structure of cold dark matter haloes}",
      journal = {\mnras},
         year = 2019,
        month = sep,
       volume = {488},
       number = {3},
        pages = {3663-3684},
          doi = {10.1093/mnras/stz1821},
archivePrefix = {arXiv},
       eprint = {1812.05777},
 primaryClass = {astro-ph.CO},
       adsurl = {https://ui.adsabs.harvard.edu/abs/2019MNRAS.488.3663L}
}

@ARTICLE{Power2003,
       author = {{Power}, C. and {Navarro}, J.~F. and {Jenkins}, A. and {Frenk}, C.~S. and {White}, S.~D.~M. and {Springel}, V. and {Stadel}, J. and {Quinn}, T.},
        title = "{The inner structure of {\ensuremath{\Lambda}}CDM haloes - I. A numerical convergence study}",
      journal = {\mnras},
         year = 2003,
        month = jan,
       volume = {338},
       number = {1},
        pages = {14-34},
          doi = {10.1046/j.1365-8711.2003.05925.x},
archivePrefix = {arXiv},
       eprint = {astro-ph/0201544},
 primaryClass = {astro-ph},
       adsurl = {https://ui.adsabs.harvard.edu/abs/2003MNRAS.338...14P}
}

@ARTICLE{Nishizawa2018,
       author = {{Nishizawa}, Atsushi J. and {Oguri}, Masamune and {Oogi}, Taira and {More}, Surhud and {Nishimichi}, Takahiro and {Nagashima}, Masahiro and {Lin}, Yen-Ting and {Mandelbaum}, Rachel and {Takada}, Masahiro and {Bahcall}, Neta and {Coupon}, Jean and {Huang}, Song and {Jian}, Hung-Yu and {Komiyama}, Yutaka and {Leauthaud}, Alexie and {Lin}, Lihwai and {Miyatake}, Hironao and {Miyazaki}, Satoshi and {Tanaka}, Masayuki},
        title = "{First results on the cluster galaxy population from the Subaru Hyper Suprime-Cam survey. II. Faint end color-magnitude diagrams and radial profiles of red and blue galaxies at 0.1 < z < 1.1}",
      journal = {\pasj},
         year = 2018,
        month = jan,
       volume = {70},
          eid = {S24},
        pages = {S24},
          doi = {10.1093/pasj/psx106},
archivePrefix = {arXiv},
       eprint = {1709.01136},
 primaryClass = {astro-ph.CO},
       adsurl = {https://ui.adsabs.harvard.edu/abs/2018PASJ...70S..24N}
}

@ARTICLE{Xu2024,
       author = {{Xu}, Weiwei and {Shan}, Huanyuan and {Li}, Ran and {Yao}, Ji and {Wang}, Chunxiang and {Li}, Nan and {Zhang}, Chaoli},
        title = "{The Measurement of the Splash-back Radius of Dark Matter Halos}",
      journal = {\apj},
         year = 2024,
        month = aug,
       volume = {971},
       number = {2},
          eid = {157},
        pages = {157},
          doi = {10.3847/1538-4357/ad57c7},
archivePrefix = {arXiv},
       eprint = {2406.06693},
 primaryClass = {astro-ph.CO},
       adsurl = {https://ui.adsabs.harvard.edu/abs/2024ApJ...971..157X}
}

@ARTICLE{Adhikari2016,
       author = {{Adhikari}, Susmita and {Dalal}, Neal and {Clampitt}, Joseph},
        title = "{Observing dynamical friction in galaxy clusters}",
      journal = {\jcap},
         year = 2016,
        month = jul,
       volume = {2016},
       number = {7},
          eid = {022},
        pages = {022},
          doi = {10.1088/1475-7516/2016/07/022},
archivePrefix = {arXiv},
       eprint = {1605.06688},
 primaryClass = {astro-ph.CO},
       adsurl = {https://ui.adsabs.harvard.edu/abs/2016JCAP...07..022A}
}

@ARTICLE{Contini2021,
       author = {{Contini}, Emanuele},
        title = "{On the Origin and Evolution of the Intra-Cluster Light: A Brief Review of the Most Recent Developments}",
      journal = {Galaxies},
         year = 2021,
        month = aug,
       volume = {9},
       number = {3},
          eid = {60},
        pages = {60},
          doi = {10.3390/galaxies9030060},
archivePrefix = {arXiv},
       eprint = {2107.04180},
 primaryClass = {astro-ph.GA},
       adsurl = {https://ui.adsabs.harvard.edu/abs/2021Galax...9...60C}
}

@ARTICLE{Montes2019,
       author = {{Montes}, Mireia and {Trujillo}, Ignacio},
        title = "{Intracluster light: a luminous tracer for dark matter in clusters of galaxies}",
      journal = {\mnras},
         year = 2019,
        month = jan,
       volume = {482},
       number = {2},
        pages = {2838-2851},
          doi = {10.1093/mnras/sty2858},
archivePrefix = {arXiv},
       eprint = {1807.11488},
 primaryClass = {astro-ph.GA},
       adsurl = {https://ui.adsabs.harvard.edu/abs/2019MNRAS.482.2838M}
}

@article{Loken_2010,
doi = {10.1088/1742-6596/256/1/012026},
url = {https://dx.doi.org/10.1088/1742-6596/256/1/012026},
year = {2010},
month = {nov},
publisher = {},
volume = {256},
number = {1},
pages = {012026},
author = {Chris Loken and Daniel Gruner and Leslie Groer and Richard Peltier and Neil Bunn and Michael Craig and Teresa Henriques and Jillian Dempsey and Ching-Hsing Yu and Joseph Chen and L Jonathan Dursi and Jason Chong and Scott Northrup and Jaime Pinto and Neil Knecht and Ramses Van Zon},
title = {SciNet: Lessons Learned from Building a Power-efficient Top-20 System and Data Centre},
journal = {Journal of Physics: Conference Series}
}

@ARTICLE{Dacunha2025,
       author = {{Dacunha}, Tara and {Mansfield}, Phil and {Wechsler}, Risa H.},
        title = "{Memoirs of Mass Accretion: Probing the Edges of Intracluster Light in Simulated Galaxy Clusters}",
      journal = {\apj},
         year = 2025,
        month = dec,
       volume = {994},
       number = {2},
          eid = {274},
        pages = {274},
          doi = {10.3847/1538-4357/ae1031},
archivePrefix = {arXiv},
       eprint = {2508.02837},
 primaryClass = {astro-ph.GA},
       adsurl = {https://ui.adsabs.harvard.edu/abs/2025ApJ...994..274D}
}

@ARTICLE{Ludlow2009,
       author = {{Ludlow}, Aaron D. and {Navarro}, Julio F. and {Springel}, Volker and {Jenkins}, Adrian and {Frenk}, Carlos S. and {Helmi}, Amina},
        title = "{The Unorthodox Orbits of Substructure Halos}",
      journal = {\apj},
         year = 2009,
        month = feb,
       volume = {692},
       number = {1},
        pages = {931-941},
          doi = {10.1088/0004-637X/692/1/931},
archivePrefix = {arXiv},
       eprint = {0801.1127},
 primaryClass = {astro-ph},
       adsurl = {https://ui.adsabs.harvard.edu/abs/2009ApJ...692..931L}
}

@ARTICLE{Bakels2021,
       author = {{Bakels}, Lucie and {Ludlow}, Aaron D. and {Power}, Chris},
        title = "{Pre-processing, group accretion, and the orbital trajectories of associated subhaloes}",
      journal = {\mnras},
         year = 2021,
        month = mar,
       volume = {501},
       number = {4},
        pages = {5948-5963},
          doi = {10.1093/mnras/staa3979},
archivePrefix = {arXiv},
       eprint = {2008.05475},
 primaryClass = {astro-ph.GA},
       adsurl = {https://ui.adsabs.harvard.edu/abs/2021MNRAS.501.5948B}
}

@ARTICLE{Merritt2005,
       author = {{Merritt}, David and {Navarro}, Julio F. and {Ludlow}, Aaron and {Jenkins}, Adrian},
        title = "{A Universal Density Profile for Dark and Luminous Matter?}",
      journal = {\apjl},
         year = 2005,
        month = may,
       volume = {624},
       number = {2},
        pages = {L85-L88},
          doi = {10.1086/430636},
archivePrefix = {arXiv},
       eprint = {astro-ph/0502515},
 primaryClass = {astro-ph},
       adsurl = {https://ui.adsabs.harvard.edu/abs/2005ApJ...624L..85M}
}

@ARTICLE{Kravtsov.Borgani.2012,
       author = {{Kravtsov}, Andrey V. and {Borgani}, Stefano},
        title = "{Formation of Galaxy Clusters}",
      journal = {\araa},
         year = 2012,
        month = sep,
       volume = {50},
        pages = {353-409},
          doi = {10.1146/annurev-astro-081811-125502},
archivePrefix = {arXiv},
       eprint = {1205.5556},
 primaryClass = {astro-ph.CO},
       adsurl = {https://ui.adsabs.harvard.edu/abs/2012ARA&A..50..353K}
}

@ARTICLE{Zhang.etal.2025,
       author = {{Zhang}, Ming and {Walker}, Kris and {Sullivan}, Andrew and {Power}, Chris and {Cui}, Weiguang and {Li}, Yichao and {Zhang}, Xin},
        title = "{The Three Hundred project: The relationship between the shock and splashback radii of simulated galaxy clusters}",
      journal = {\pasa},
         year = 2025,
        month = jan,
       volume = {42},
          eid = {e008},
        pages = {e008},
          doi = {10.1017/pasa.2024.132},
archivePrefix = {arXiv},
       eprint = {2412.09864},
 primaryClass = {astro-ph.CO},
       adsurl = {https://ui.adsabs.harvard.edu/abs/2025PASA...42....8Z}
}

@ARTICLE{Gabriel-Silve.Sodre.2025,
       author = {{Gabriel-Silva}, Lucas and {Sodr{\'e}}, Jr., Laerte},
        title = "{Galaxy Cluster Mass Estimation through the Splashback Radius}",
      journal = {\apj},
         year = 2025,
        month = aug,
       volume = {988},
       number = {2},
          eid = {149},
        pages = {149},
          doi = {10.3847/1538-4357/ade308},
archivePrefix = {arXiv},
       eprint = {2506.07425},
 primaryClass = {astro-ph.CO},
       adsurl = {https://ui.adsabs.harvard.edu/abs/2025ApJ...988..149G}
}

@ARTICLE{White.2001,
       author = {{White}, M.},
        title = "{The mass of a halo}",
      journal = {\aap},
         year = 2001,
        month = feb,
       volume = {367},
        pages = {27-32},
          doi = {10.1051/0004-6361:20000357},
archivePrefix = {arXiv},
       eprint = {astro-ph/0011495},
 primaryClass = {astro-ph},
       adsurl = {https://ui.adsabs.harvard.edu/abs/2001A&A...367...27W}
}

@ARTICLE{Lukic.etal.2009,
       author = {{Luki{\'c}}, Zarija and {Reed}, Darren and {Habib}, Salman and {Heitmann}, Katrin},
        title = "{The Structure of Halos: Implications for Group and Cluster Cosmology}",
      journal = {\apj},
         year = 2009,
        month = feb,
       volume = {692},
       number = {1},
        pages = {217-228},
          doi = {10.1088/0004-637X/692/1/217},
archivePrefix = {arXiv},
       eprint = {0803.3624},
 primaryClass = {astro-ph},
       adsurl = {https://ui.adsabs.harvard.edu/abs/2009ApJ...692..217L}
}

@ARTICLE{Morandi.Sun.2016,
       author = {{Morandi}, Andrea and {Sun}, Ming},
        title = "{Probing dark energy via galaxy cluster outskirts}",
      journal = {\mnras},
         year = 2016,
        month = apr,
       volume = {457},
       number = {3},
        pages = {3266-3284},
          doi = {10.1093/mnras/stw143},
archivePrefix = {arXiv},
       eprint = {1601.03741},
 primaryClass = {astro-ph.CO},
       adsurl = {https://ui.adsabs.harvard.edu/abs/2016MNRAS.457.3266M}
}

@ARTICLE{Haggar.etal.2024,
       author = {{Haggar}, Roan and {Amoura}, Yuba and {Mpetha}, Charlie T. and {Taylor}, James E. and {Walker}, Kris and {Power}, Chris},
        title = "{Constraining Cosmological Parameters Using the Splashback Radius of Galaxy Clusters}",
      journal = {\apj},
         year = 2024,
        month = sep,
       volume = {972},
       number = {1},
          eid = {28},
        pages = {28},
          doi = {10.3847/1538-4357/ad5cee},
archivePrefix = {arXiv},
       eprint = {2406.17849},
 primaryClass = {astro-ph.CO},
       adsurl = {https://ui.adsabs.harvard.edu/abs/2024ApJ...972...28H}
}

@ARTICLE{Eke.etal.2004,
       author = {{Eke}, V.~R. and {Baugh}, Carlton M. and {Cole}, Shaun and {Frenk}, Carlos S. and {Norberg}, Peder and {Peacock}, John A. and {Baldry}, Ivan K. and {Bland-Hawthorn}, Joss and {Bridges}, Terry and {Cannon}, Russell and {Colless}, Matthew and {Collins}, Chris and {Couch}, Warrick and {Dalton}, Gavin and {de Propris}, Roberto and {Driver}, Simon P. and {Efstathiou}, George and {Ellis}, Richard S. and {Glazebrook}, Karl and {Jackson}, Carole and {Lahav}, Ofer and {Lewis}, Ian and {Lumsden}, Stuart and {Maddox}, Steve and {Madgwick}, Darren and {Peterson}, Bruce A. and {Sutherland}, Will and {Taylor}, Keith},
        title = "{Galaxy groups in the 2dFGRS: the group-finding algorithm and the 2PIGG catalogue}",
      journal = {\mnras},
         year = 2004,
        month = mar,
       volume = {348},
       number = {3},
        pages = {866-878},
          doi = {10.1111/j.1365-2966.2004.07408.x},
archivePrefix = {arXiv},
       eprint = {astro-ph/0402567},
 primaryClass = {astro-ph},
       adsurl = {https://ui.adsabs.harvard.edu/abs/2004MNRAS.348..866E}
}

@ARTICLE{Old.etal.2014,
       author = {{Old}, L. and {Skibba}, R.~A. and {Pearce}, F.~R. and {Croton}, D. and {Muldrew}, S.~I. and {Mu{\~n}oz-Cuartas}, J.~C. and {Gifford}, D. and {Gray}, M.~E. and {von der Linden}, A. and {Mamon}, G.~A. and {Merrifield}, M.~R. and {M{\"u}ller}, V. and {Pearson}, R.~J. and {Ponman}, T.~J. and {Saro}, A. and {Sepp}, T. and {Sif{\'o}n}, C. and {Tempel}, E. and {Tundo}, E. and {Wang}, Y.~O. and {Wojtak}, R.},
        title = "{Galaxy cluster mass reconstruction project - I. Methods and first results on galaxy-based techniques}",
      journal = {\mnras},
         year = 2014,
        month = jun,
       volume = {441},
       number = {2},
        pages = {1513-1536},
          doi = {10.1093/mnras/stu545},
archivePrefix = {arXiv},
       eprint = {1403.4610},
 primaryClass = {astro-ph.CO},
       adsurl = {https://ui.adsabs.harvard.edu/abs/2014MNRAS.441.1513O}
}

@ARTICLE{Old.etal.2015,
       author = {{Old}, L. and {Wojtak}, R. and {Mamon}, G.~A. and {Skibba}, R.~A. and {Pearce}, F.~R. and {Croton}, D. and {Bamford}, S. and {Behroozi}, P. and {de Carvalho}, R. and {Mu{\~n}oz-Cuartas}, J.~C. and {Gifford}, D. and {Gray}, M.~E. and {von der Linden}, A. and {Merrifield}, M.~R. and {Muldrew}, S.~I. and {M{\"u}ller}, V. and {Pearson}, R.~J. and {Ponman}, T.~J. and {Rozo}, E. and {Rykoff}, E. and {Saro}, A. and {Sepp}, T. and {Sif{\'o}n}, C. and {Tempel}, E.},
        title = "{Galaxy Cluster Mass Reconstruction Project - II. Quantifying scatter and bias using contrasting mock catalogues}",
      journal = {\mnras},
         year = 2015,
        month = may,
       volume = {449},
       number = {2},
        pages = {1897-1920},
          doi = {10.1093/mnras/stv421},
archivePrefix = {arXiv},
       eprint = {1502.07347},
 primaryClass = {astro-ph.CO},
       adsurl = {https://ui.adsabs.harvard.edu/abs/2015MNRAS.449.1897O}
}

@ARTICLE{Lebeau.etal.2024,
       author = {{Lebeau}, Th{\'e}o and {Ettori}, Stefano and {Aghanim}, Nabila and {Sorce}, Jenny G.},
        title = "{Can the splashback radius be an observable boundary of galaxy clusters?}",
      journal = {\aap},
         year = 2024,
        month = sep,
       volume = {689},
          eid = {A19},
        pages = {A19},
          doi = {10.1051/0004-6361/202450146},
archivePrefix = {arXiv},
       eprint = {2403.18648},
 primaryClass = {astro-ph.CO},
       adsurl = {https://ui.adsabs.harvard.edu/abs/2024A&A...689A..19L}
}

@ARTICLE{Aung.etal.2021,
       author = {{Aung}, Han and {Nagai}, Daisuke and {Rozo}, Eduardo and {Garc{\'\i}a}, Rafael},
        title = "{The phase-space structure of dark matter haloes}",
      journal = {\mnras},
         year = 2021,
        month = mar,
       volume = {502},
       number = {1},
        pages = {1041-1047},
          doi = {10.1093/mnras/staa3994},
archivePrefix = {arXiv},
       eprint = {2003.11557},
 primaryClass = {astro-ph.CO},
       adsurl = {https://ui.adsabs.harvard.edu/abs/2021MNRAS.502.1041A}
}

@ARTICLE{ONeil.etal.2022,
       author = {{O'Neil}, Stephanie and {Borrow}, Josh and {Vogelsberger}, Mark and {Diemer}, Benedikt},
        title = "{The impact of galaxy selection on the splashback boundaries of galaxy clusters}",
      journal = {\mnras},
         year = 2022,
        month = jun,
       volume = {513},
       number = {1},
        pages = {835-852},
          doi = {10.1093/mnras/stac850},
archivePrefix = {arXiv},
       eprint = {2202.05277},
 primaryClass = {astro-ph.GA},
       adsurl = {https://ui.adsabs.harvard.edu/abs/2022MNRAS.513..835O}
}

@ARTICLE{Brown.etal.2024,
       author = {{Brown}, Harley J. and {Martin}, Garreth and {Pearce}, Frazer R. and {Hatch}, Nina A. and {Bah{\'e}}, Yannick M. and {Dubois}, Yohan},
        title = "{Assembly of the intracluster light in the HORIZON-AGN simulation}",
      journal = {\mnras},
         year = 2024,
        month = oct,
       volume = {534},
       number = {1},
        pages = {431-443},
          doi = {10.1093/mnras/stae2084},
archivePrefix = {arXiv},
       eprint = {2409.10607},
 primaryClass = {astro-ph.GA},
       adsurl = {https://ui.adsabs.harvard.edu/abs/2024MNRAS.534..431B}
}

@ARTICLE{Towler.etal.2024,
       author = {{Towler}, Imogen and {Kay}, Scott T. and {Schaye}, Joop and {Kugel}, Roi and {Schaller}, Matthieu and {Braspenning}, Joey and {Elbers}, Willem and {Frenk}, Carlos S. and {Kwan}, Juliana and {Salcido}, Jaime and {van Daalen}, Marcel P. and {Vandenbroucke}, Bert and {Altamura}, Edoardo},
        title = "{Inferring the dark matter splashback radius from cluster gas and observable profiles in the FLAMINGO simulations}",
      journal = {\mnras},
         year = 2024,
        month = apr,
       volume = {529},
       number = {3},
        pages = {2017-2031},
          doi = {10.1093/mnras/stae654},
archivePrefix = {arXiv},
       eprint = {2312.05126},
 primaryClass = {astro-ph.CO},
       adsurl = {https://ui.adsabs.harvard.edu/abs/2024MNRAS.529.2017T}
}

@ARTICLE{Kluge.etal.2025,
       author = {{Kluge}, M. and {Hatch}, N.~A. and {Montes}, M. and {Golden-Marx}, J.~B. and {Gonzalez}, A.~H. and {Cuillandre}, J. -C. and {Bolzonella}, M. and {Lan{\c{c}}on}, A. and {Laureijs}, R. and {Saifollahi}, T. and {Schirmer}, M. and {Stone}, C. and {Boselli}, A. and {Cantiello}, M. and {Sorce}, J.~G. and {Marleau}, F.~R. and {Duc}, P. -A. and {Sola}, E. and {Urbano}, M. and {Ahad}, S.~L. and {Bah{\'e}}, Y.~M. and {Bamford}, S.~P. and {Bellhouse}, C. and {Buitrago}, F. and {Dimauro}, P. and {Durret}, F. and {Ellien}, A. and {Jimenez-Teja}, Y. and {Slezak}, E. and {Aghanim}, N. and {Altieri}, B. and {Andreon}, S. and {Auricchio}, N. and {Baldi}, M. and {Balestra}, A. and {Bardelli}, S. and {Bender}, R. and {Bonino}, D. and {Branchini}, E. and {Brescia}, M. and {Brinchmann}, J. and {Camera}, S. and {Candini}, G.~P. and {Capobianco}, V. and {Carbone}, C. and {Carretero}, J. and {Casas}, S. and {Castellano}, M. and {Cavuoti}, S. and {Cimatti}, A. and {Congedo}, G. and {Conselice}, C.~J. and {Conversi}, L. and {Copin}, Y. and {Courbin}, F. and {Courtois}, H.~M. and {Cropper}, M. and {Da Silva}, A. and {Degaudenzi}, H. and {Dinis}, J. and {Duncan}, C.~A.~J. and {Dupac}, X. and {Dusini}, S. and {Farina}, M. and {Farrens}, S. and {Ferriol}, S. and {Fosalba}, P. and {Frailis}, M. and {Franceschi}, E. and {Fumana}, M. and {Galeotta}, S. and {Garilli}, B. and {Gillard}, W. and {Gillis}, B. and {Giocoli}, C. and {G{\'o}mez-Alvarez}, P. and {Granett}, B.~R. and {Grazian}, A. and {Grupp}, F. and {Guzzo}, L. and {Haugan}, S.~V.~H. and {Hoar}, J. and {Hoekstra}, H. and {Holmes}, W. and {Hook}, I. and {Hormuth}, F. and {Hornstrup}, A. and {Hudelot}, P. and {Jahnke}, K. and {Keih{\"a}nen}, E. and {Kermiche}, S. and {Kiessling}, A. and {Kitching}, T. and {Kohley}, R. and {Kubik}, B. and {K{\"u}mmel}, M. and {Kunz}, M. and {Kurki-Suonio}, H. and {Lahav}, O. and {Ligori}, S. and {Lilje}, P.~B. and {Lindholm}, V. and {Lloro}, I. and {Maiorano}, E. and {Mansutti}, O. and {Marggraf}, O. and {Markovic}, K. and {Martinet}, N. and {Marulli}, F. and {Massey}, R. and {Maurogordato}, S. and {McCracken}, H.~J. and {Medinaceli}, E. and {Mei}, S. and {Melchior}, M. and {Mellier}, Y. and {Meneghetti}, M. and {Merlin}, E. and {Meylan}, G. and {Moresco}, M. and {Moscardini}, L. and {Munari}, E. and {Nichol}, R.~C. and {Niemi}, S. -M. and {Nightingale}, J.~W. and {Padilla}, C. and {Paltani}, S. and {Pasian}, F. and {Pedersen}, K. and {Percival}, W.~J. and {Pettorino}, V. and {Pires}, S. and {Polenta}, G. and {Poncet}, M. and {Popa}, L.~A. and {Pozzetti}, L. and {Racca}, G.~D. and {Raison}, F. and {Rebolo}, R. and {Renzi}, A. and {Rhodes}, J. and {Riccio}, G. and {Rix}, H. -W. and {Romelli}, E. and {Roncarelli}, M. and {Rossetti}, E. and {Saglia}, R. and {Sapone}, D. and {Sartoris}, B. and {Sauvage}, M. and {Scaramella}, R. and {Schneider}, P. and {Schrabback}, T. and {Secroun}, A. and {Seidel}, G. and {Seiffert}, M. and {Serrano}, S. and {Sirignano}, C. and {Sirri}, G. and {Skottfelt}, J. and {Stanco}, L. and {Tallada-Cresp{\'\i}}, P. and {Taylor}, A.~N. and {Teplitz}, H.~I. and {Tereno}, I. and {Toledo-Moreo}, R. and {Torradeflot}, F. and {Tutusaus}, I. and {Valentijn}, E.~A. and {Valenziano}, L. and {Vassallo}, T. and {Verdoes Kleijn}, G. and {Veropalumbo}, A. and {Wang}, Y. and {Weller}, J. and {Williams}, O.~R. and {Zamorani}, G. and {Zucca}, E. and {Biviano}, A. and {Burigana}, C. and {De Lucia}, G. and {George}, K. and {Scottez}, V. and {Simon}, P. and {Mora}, A. and {Mart{\'\i}n-Fleitas}, J. and {Ruppin}, F. and {Scott}, D.},
        title = "{Euclid: Early Release Observations {\textendash} The intracluster light and intracluster globular clusters of the Perseus cluster}",
      journal = {\aap},
         year = 2025,
        month = may,
       volume = {697},
          eid = {A13},
        pages = {A13},
          doi = {10.1051/0004-6361/202450772},
archivePrefix = {arXiv},
       eprint = {2405.13503},
 primaryClass = {astro-ph.GA},
       adsurl = {https://ui.adsabs.harvard.edu/abs/2025A&A...697A..13K}
}

@ARTICLE{Montes.Trujillo.2018,
       author = {{Montes}, Mireia and {Trujillo}, Ignacio},
        title = "{Intracluster light at the Frontier - II. The Frontier Fields Clusters}",
      journal = {\mnras},
         year = 2018,
        month = feb,
       volume = {474},
       number = {1},
        pages = {917-932},
          doi = {10.1093/mnras/stx2847},
archivePrefix = {arXiv},
       eprint = {1710.03240},
 primaryClass = {astro-ph.CO},
       adsurl = {https://ui.adsabs.harvard.edu/abs/2018MNRAS.474..917M}
}

@ARTICLE{Caminha.etal.2017,
       author = {{Caminha}, G.~B. and {Grillo}, C. and {Rosati}, P. and {Meneghetti}, M. and {Mercurio}, A. and {Ettori}, S. and {Balestra}, I. and {Biviano}, A. and {Umetsu}, K. and {Vanzella}, E. and {Annunziatella}, M. and {Bonamigo}, M. and {Delgado-Correal}, C. and {Girardi}, M. and {Lombardi}, M. and {Nonino}, M. and {Sartoris}, B. and {Tozzi}, P. and {Bartelmann}, M. and {Bradley}, L. and {Caputi}, K.~I. and {Coe}, D. and {Ford}, H. and {Fritz}, A. and {Gobat}, R. and {Postman}, M. and {Seitz}, S. and {Zitrin}, A.},
        title = "{Mass distribution in the core of MACS J1206. Robust modeling from an exceptionally large sample of central multiple images}",
      journal = {\aap},
         year = 2017,
        month = nov,
       volume = {607},
          eid = {A93},
        pages = {A93},
          doi = {10.1051/0004-6361/201731498},
archivePrefix = {arXiv},
       eprint = {1707.00690},
 primaryClass = {astro-ph.GA},
       adsurl = {https://ui.adsabs.harvard.edu/abs/2017A&A...607A..93C}
}

@ARTICLE{Sun.etal.2025,
       author = {{Sun}, Xiaoqing and {O'Neil}, Stephanie and {Shen}, Xuejian and {Vogelsberger}, Mark},
        title = "{The effects of projection on measuring the splashback feature}",
      journal = {The Open Journal of Astrophysics},
         year = 2025,
        month = jul,
       volume = {8},
          eid = {100},
        pages = {100},
          doi = {10.33232/001c.142495},
archivePrefix = {arXiv},
       eprint = {2503.04882},
 primaryClass = {astro-ph.GA},
       adsurl = {https://ui.adsabs.harvard.edu/abs/2025OJAp....8E.100S}
}

@ARTICLE{ONeil.etal.2021,
       author = {{O'Neil}, Stephanie and {Barnes}, David J. and {Vogelsberger}, Mark and {Diemer}, Benedikt},
        title = "{The splashback boundary of haloes in hydrodynamic simulations}",
      journal = {\mnras},
         year = 2021,
        month = jul,
       volume = {504},
       number = {3},
        pages = {4649-4666},
          doi = {10.1093/mnras/stab1221},
archivePrefix = {arXiv},
       eprint = {2012.00025},
 primaryClass = {astro-ph.GA},
       adsurl = {https://ui.adsabs.harvard.edu/abs/2021MNRAS.504.4649O}
}

@ARTICLE{Pizzardo.etal.2024,
       author = {{Pizzardo}, Michele and {Geller}, Margaret J. and {Kenyon}, Scott J. and {Damjanov}, Ivana},
        title = "{The splashback radius and the radial velocity profile of galaxy clusters in IllustrisTNG}",
      journal = {\aap},
         year = 2024,
        month = mar,
       volume = {683},
          eid = {A82},
        pages = {A82},
          doi = {10.1051/0004-6361/202348643},
archivePrefix = {arXiv},
       eprint = {2311.10854},
 primaryClass = {astro-ph.CO},
       adsurl = {https://ui.adsabs.harvard.edu/abs/2024A&A...683A..82P}
}

@ARTICLE{Murata2020,
       author = {{Murata}, Ryoma and {Sunayama}, Tomomi and {Oguri}, Masamune and {More}, Surhud and {Nishizawa}, Atsushi J. and {Nishimichi}, Takahiro and {Osato}, Ken},
        title = "{The splashback radius of optically selected clusters with Subaru HSC Second Public Data Release}",
      journal = {\pasj},
         year = 2020,
        month = aug,
       volume = {72},
       number = {4},
          eid = {64},
        pages = {64},
          doi = {10.1093/pasj/psaa041},
archivePrefix = {arXiv},
       eprint = {2001.01160},
 primaryClass = {astro-ph.CO},
       adsurl = {https://ui.adsabs.harvard.edu/abs/2020PASJ...72...64M}
}

@ARTICLE{Gonzalez2021,
       author = {{Gonzalez}, Anthony H. and {George}, Tyler and {Connor}, Thomas and {Deason}, Alis and {Donahue}, Megan and {Montes}, Mireia and {Zabludoff}, Ann I. and {Zaritsky}, Dennis},
        title = "{Discovery of a possible splashback feature in the intracluster light of MACS J1149.5+2223}",
      journal = {\mnras},
         year = 2021,
        month = oct,
       volume = {507},
       number = {1},
        pages = {963-970},
          doi = {10.1093/mnras/stab2117},
archivePrefix = {arXiv},
       eprint = {2104.04306},
 primaryClass = {astro-ph.CO},
       adsurl = {https://ui.adsabs.harvard.edu/abs/2021MNRAS.507..963G}
}

@ARTICLE{Rana2023,
       author = {{Rana}, Divya and {More}, Surhud and {Miyatake}, Hironao and {Grandis}, Sebastian and {Klein}, Matthias and {Bulbul}, Esra and {Chiu}, I. -Non and {Miyazaki}, Satoshi and {Bahcall}, Neta},
        title = "{The eROSITA Final Equatorial-Depth Survey (eFEDS) - Splashback radius of X-ray galaxy clusters using galaxies from HSC survey}",
      journal = {\mnras},
         year = 2023,
        month = jul,
       volume = {522},
       number = {3},
        pages = {4181-4195},
          doi = {10.1093/mnras/stad1239},
archivePrefix = {arXiv},
       eprint = {2301.03626},
 primaryClass = {astro-ph.CO},
       adsurl = {https://ui.adsabs.harvard.edu/abs/2023MNRAS.522.4181R}
}

@ARTICLE{Busch2017,
       author = {{Busch}, Philipp and {White}, Simon D.~M.},
        title = "{Assembly bias and splashback in galaxy clusters}",
      journal = {\mnras},
         year = 2017,
        month = oct,
       volume = {470},
       number = {4},
        pages = {4767-4781},
          doi = {10.1093/mnras/stx1584},
archivePrefix = {arXiv},
       eprint = {1702.01682},
 primaryClass = {astro-ph.CO},
       adsurl = {https://ui.adsabs.harvard.edu/abs/2017MNRAS.470.4767B}
}

@ARTICLE{Sifon2018,
       author = {{Sif{\'o}n}, Crist{\'o}bal and {Herbonnet}, Ricardo and {Hoekstra}, Henk and {van der Burg}, Remco F.~J. and {Viola}, Massimo},
        title = "{The galaxy-subhalo connection in low-redshift galaxy clusters from weak gravitational lensing}",
      journal = {\mnras},
         year = 2018,
        month = jul,
       volume = {478},
       number = {1},
        pages = {1244-1264},
          doi = {10.1093/mnras/sty1161},
archivePrefix = {arXiv},
       eprint = {1706.06125},
 primaryClass = {astro-ph.GA},
       adsurl = {https://ui.adsabs.harvard.edu/abs/2018MNRAS.478.1244S}
}

@ARTICLE{Kumar2026,
       author = {{Kumar}, Amit and {More}, Surhud},
        title = "{Environmental dependence of the galaxy─halo connection of satellites using HSC weak lensing}",
      journal = {\mnras},
         year = 2026,
        month = feb,
       volume = {545},
       number = {4},
          eid = {staf2026},
        pages = {staf2026},
          doi = {10.1093/mnras/staf2026},
archivePrefix = {arXiv},
       eprint = {2409.05795},
 primaryClass = {astro-ph.CO},
       adsurl = {https://ui.adsabs.harvard.edu/abs/2026MNRAS.545f2026K}
}

@ARTICLE{ContrerasSantos2024,
       author = {{Contreras-Santos}, A. and {Knebe}, A. and {Cui}, W. and {Alonso Asensio}, I. and {Dalla Vecchia}, C. and {Ca{\~n}as}, R. and {Haggar}, R. and {Mostoghiu Paun}, R.~A. and {Pearce}, F.~R. and {Rasia}, E.},
        title = "{Characterising the intra-cluster light in The Three Hundred simulations}",
      journal = {Astronomy and Astrophysics},
         year = 2024,
        month = mar,
       volume = {683},
          eid = {A59},
        pages = {A59},
          doi = {10.1051/0004-6361/202348474},
archivePrefix = {arXiv},
       eprint = {2401.08283},
 primaryClass = {astro-ph.CO},
       adsurl = {https://ui.adsabs.harvard.edu/abs/2024A&A...683A..59C}
}

@ARTICLE{Giocoli2024,
       author = {{Giocoli}, Carlo and {Palmucci}, Lorenzo and {Lesci}, Giorgio F. and {Moscardini}, Lauro and {Despali}, Giulia and {Marulli}, Federico and {Maturi}, Matteo and {Radovich}, Mario and {Sereno}, Mauro and {Bardelli}, Sandro and {Castignani}, Gianluca and {Covone}, Giovanni and {Ingoglia}, Lorenzo and {Romanello}, Massimiliano and {Roncarelli}, Mauro and {Puddu}, Emanuella},
        title = "{AMICO galaxy clusters in KiDS-DR3: Measuring the splashback radius from weak gravitational lensing}",
      journal = {\aap},
         year = 2024,
        month = jul,
       volume = {687},
          eid = {A79},
        pages = {A79},
          doi = {10.1051/0004-6361/202449561},
archivePrefix = {arXiv},
       eprint = {2402.06717},
 primaryClass = {astro-ph.CO},
       adsurl = {https://ui.adsabs.harvard.edu/abs/2024A&A...687A..79G}
}

@ARTICLE{Gill2004,
       author = {{Gill}, Stuart P.~D. and {Knebe}, Alexander and {Gibson}, Brad K.},
        title = "{The evolution of substructure - I. A new identification method}",
      journal = {\mnras},
         year = 2004,
        month = jun,
       volume = {351},
       number = {2},
        pages = {399-409},
          doi = {10.1111/j.1365-2966.2004.07786.x},
archivePrefix = {arXiv},
       eprint = {astro-ph/0404258},
 primaryClass = {astro-ph},
       adsurl = {https://ui.adsabs.harvard.edu/abs/2004MNRAS.351..399G}
}

@ARTICLE{Zhang2019,
       author = {{Zhang}, Y. and {Yanny}, B. and {Palmese}, A. and {Gruen}, D. and {To}, C. and {Rykoff}, E.~S. and {Leung}, Y. and {Collins}, C. and {Hilton}, M. and {Abbott}, T.~M.~C. and {Annis}, J. and {Avila}, S. and {Bertin}, E. and {Brooks}, D. and {Burke}, D.~L. and {Carnero Rosell}, A. and {Carrasco Kind}, M. and {Carretero}, J. and {Cunha}, C.~E. and {D'Andrea}, C.~B. and {da Costa}, L.~N. and {De Vicente}, J. and {Desai}, S. and {Diehl}, H.~T. and {Dietrich}, J.~P. and {Doel}, P. and {Drlica-Wagner}, A. and {Eifler}, T.~F. and {Evrard}, A.~E. and {Flaugher}, B. and {Fosalba}, P. and {Frieman}, J. and {Garc{\'\i}a-Bellido}, J. and {Gaztanaga}, E. and {Gerdes}, D.~W. and {Gruendl}, R.~A. and {Gschwend}, J. and {Gutierrez}, G. and {Hartley}, W.~G. and {Hollowood}, D.~L. and {Honscheid}, K. and {Hoyle}, B. and {James}, D.~J. and {Jeltema}, T. and {Kuehn}, K. and {Kuropatkin}, N. and {Li}, T.~S. and {Lima}, M. and {Maia}, M.~A.~G. and {March}, M. and {Marshall}, J.~L. and {Melchior}, P. and {Menanteau}, F. and {Miller}, C.~J. and {Miquel}, R. and {Mohr}, J.~J. and {Ogando}, R.~L.~C. and {Plazas}, A.~A. and {Romer}, A.~K. and {Sanchez}, E. and {Scarpine}, V. and {Schubnell}, M. and {Serrano}, S. and {Sevilla-Noarbe}, I. and {Smith}, M. and {Soares-Santos}, M. and {Sobreira}, F. and {Suchyta}, E. and {Swanson}, M.~E.~C. and {Tarle}, G. and {Thomas}, D. and {Wester}, W. and {DES Collaboration}},
        title = "{Dark Energy Survey Year 1 Results: Detection of Intracluster Light at Redshift {\ensuremath{\sim}} 0.25}",
      journal = {\apj},
         year = 2019,
        month = apr,
       volume = {874},
       number = {2},
          eid = {165},
        pages = {165},
          doi = {10.3847/1538-4357/ab0dfd},
archivePrefix = {arXiv},
       eprint = {1812.04004},
 primaryClass = {astro-ph.CO},
       adsurl = {https://ui.adsabs.harvard.edu/abs/2019ApJ...874..165Z}
}

@ARTICLE{Barnes2017,
       author = {{Barnes}, David J. and {Kay}, Scott T. and {Bah{\'e}}, Yannick M. and {Dalla Vecchia}, Claudio and {McCarthy}, Ian G. and {Schaye}, Joop and {Bower}, Richard G. and {Jenkins}, Adrian and {Thomas}, Peter A. and {Schaller}, Matthieu and {Crain}, Robert A. and {Theuns}, Tom and {White}, Simon D.~M.},
        title = "{The Cluster-EAGLE project: global properties of simulated clusters with resolved galaxies}",
      journal = {\mnras},
         year = 2017,
        month = oct,
       volume = {471},
       number = {1},
        pages = {1088-1106},
          doi = {10.1093/mnras/stx1647},
archivePrefix = {arXiv},
       eprint = {1703.10907},
 primaryClass = {astro-ph.GA},
       adsurl = {https://ui.adsabs.harvard.edu/abs/2017MNRAS.471.1088B}
}

@ARTICLE{Ivezic2019,
       author = {{Ivezi{\'c}}, {\v{Z}}eljko and {Kahn}, Steven M. and {Tyson}, J. Anthony and {Abel}, Bob and {Acosta}, Emily and {Allsman}, Robyn and {Alonso}, David and {AlSayyad}, Yusra and {Anderson}, Scott F. and {Andrew}, John and {Angel}, James Roger P. and {Angeli}, George Z. and {Ansari}, Reza and {Antilogus}, Pierre and {Araujo}, Constanza and {Armstrong}, Robert and {Arndt}, Kirk T. and {Astier}, Pierre and {Aubourg}, {\'E}ric and {Auza}, Nicole and {Axelrod}, Tim S. and {Bard}, Deborah J. and {Barr}, Jeff D. and {Barrau}, Aurelian and {Bartlett}, James G. and {Bauer}, Amanda E. and {Bauman}, Brian J. and {Baumont}, Sylvain and {Bechtol}, Ellen and {Bechtol}, Keith and {Becker}, Andrew C. and {Becla}, Jacek and {Beldica}, Cristina and {Bellavia}, Steve and {Bianco}, Federica B. and {Biswas}, Rahul and {Blanc}, Guillaume and {Blazek}, Jonathan and {Blandford}, Roger D. and {Bloom}, Josh S. and {Bogart}, Joanne and {Bond}, Tim W. and {Booth}, Michael T. and {Borgland}, Anders W. and {Borne}, Kirk and {Bosch}, James F. and {Boutigny}, Dominique and {Brackett}, Craig A. and {Bradshaw}, Andrew and {Brandt}, William Nielsen and {Brown}, Michael E. and {Bullock}, James S. and {Burchat}, Patricia and {Burke}, David L. and {Cagnoli}, Gianpietro and {Calabrese}, Daniel and {Callahan}, Shawn and {Callen}, Alice L. and {Carlin}, Jeffrey L. and {Carlson}, Erin L. and {Chandrasekharan}, Srinivasan and {Charles-Emerson}, Glenaver and {Chesley}, Steve and {Cheu}, Elliott C. and {Chiang}, Hsin-Fang and {Chiang}, James and {Chirino}, Carol and {Chow}, Derek and {Ciardi}, David R. and {Claver}, Charles F. and {Cohen-Tanugi}, Johann and {Cockrum}, Joseph J. and {Coles}, Rebecca and {Connolly}, Andrew J. and {Cook}, Kem H. and {Cooray}, Asantha and {Covey}, Kevin R. and {Cribbs}, Chris and {Cui}, Wei and {Cutri}, Roc and {Daly}, Philip N. and {Daniel}, Scott F. and {Daruich}, Felipe and {Daubard}, Guillaume and {Daues}, Greg and {Dawson}, William and {Delgado}, Francisco and {Dellapenna}, Alfred and {de Peyster}, Robert and {de Val-Borro}, Miguel and {Digel}, Seth W. and {Doherty}, Peter and {Dubois}, Richard and {Dubois-Felsmann}, Gregory P. and {Durech}, Josef and {Economou}, Frossie and {Eifler}, Tim and {Eracleous}, Michael and {Emmons}, Benjamin L. and {Fausti Neto}, Angelo and {Ferguson}, Henry and {Figueroa}, Enrique and {Fisher-Levine}, Merlin and {Focke}, Warren and {Foss}, Michael D. and {Frank}, James and {Freemon}, Michael D. and {Gangler}, Emmanuel and {Gawiser}, Eric and {Geary}, John C. and {Gee}, Perry and {Geha}, Marla and {Gessner}, Charles J.~B. and {Gibson}, Robert R. and {Gilmore}, D. Kirk and {Glanzman}, Thomas and {Glick}, William and {Goldina}, Tatiana and {Goldstein}, Daniel A. and {Goodenow}, Iain and {Graham}, Melissa L. and {Gressler}, William J. and {Gris}, Philippe and {Guy}, Leanne P. and {Guyonnet}, Augustin and {Haller}, Gunther and {Harris}, Ron and {Hascall}, Patrick A. and {Haupt}, Justine and {Hernandez}, Fabio and {Herrmann}, Sven and {Hileman}, Edward and {Hoblitt}, Joshua and {Hodgson}, John A. and {Hogan}, Craig and {Howard}, James D. and {Huang}, Dajun and {Huffer}, Michael E. and {Ingraham}, Patrick and {Innes}, Walter R. and {Jacoby}, Suzanne H. and {Jain}, Bhuvnesh and {Jammes}, Fabrice and {Jee}, M. James and {Jenness}, Tim and {Jernigan}, Garrett and {Jevremovi{\'c}}, Darko and {Johns}, Kenneth and {Johnson}, Anthony S. and {Johnson}, Margaret W.~G. and {Jones}, R. Lynne and {Juramy-Gilles}, Claire and {Juri{\'c}}, Mario and {Kalirai}, Jason S. and {Kallivayalil}, Nitya J. and {Kalmbach}, Bryce and {Kantor}, Jeffrey P. and {Karst}, Pierre and {Kasliwal}, Mansi M. and {Kelly}, Heather and {Kessler}, Richard and {Kinnison}, Veronica and {Kirkby}, David and {Knox}, Lloyd and {Kotov}, Ivan V. and {Krabbendam}, Victor L. and {Krughoff}, K. Simon and {Kub{\'a}nek}, Petr and {Kuczewski}, John and {Kulkarni}, Shri and {Ku}, John and {Kurita}, Nadine R. and {Lage}, Craig S. and {Lambert}, Ron and {Lange}, Travis and {Langton}, J. Brian and {Le Guillou}, Laurent and {Levine}, Deborah and {Liang}, Ming and {Lim}, Kian-Tat and {Lintott}, Chris J. and {Long}, Kevin E. and {Lopez}, Margaux and {Lotz}, Paul J. and {Lupton}, Robert H. and {Lust}, Nate B. and {MacArthur}, Lauren A. and {Mahabal}, Ashish and {Mandelbaum}, Rachel and {Markiewicz}, Thomas W. and {Marsh}, Darren S. and {Marshall}, Philip J. and {Marshall}, Stuart and {May}, Morgan and {McKercher}, Robert and {McQueen}, Michelle and {Meyers}, Joshua and {Migliore}, Myriam and {Miller}, Michelle and {Mills}, David J.},
        title = "{LSST: From Science Drivers to Reference Design and Anticipated Data Products}",
      journal = {\apj},
         year = 2019,
        month = mar,
       volume = {873},
       number = {2},
          eid = {111},
        pages = {111},
          doi = {10.3847/1538-4357/ab042c},
archivePrefix = {arXiv},
       eprint = {0805.2366},
 primaryClass = {astro-ph},
       adsurl = {https://ui.adsabs.harvard.edu/abs/2019ApJ...873..111I}
}

@ARTICLE{Xie2015,
       author = {{Xie}, Lizhi and {Gao}, Liang},
        title = "{Assembly history of subhalo populations in galactic and cluster sized dark haloes}",
      journal = {\mnras},
         year = 2015,
        month = dec,
       volume = {454},
       number = {2},
        pages = {1697-1703},
          doi = {10.1093/mnras/stv2077},
archivePrefix = {arXiv},
       eprint = {1501.03171},
 primaryClass = {astro-ph.CO},
       adsurl = {https://ui.adsabs.harvard.edu/abs/2015MNRAS.454.1697X}
}

@ARTICLE{Sunayama2020,
       author = {{Sunayama}, Tomomi and {Park}, Youngsoo and {Takada}, Masahiro and {Kobayashi}, Yosuke and {Nishimichi}, Takahiro and {Kurita}, Toshiki and {More}, Surhud and {Oguri}, Masamune and {Osato}, Ken},
        title = "{The impact of projection effects on cluster observables: stacked lensing and projected clustering}",
      journal = {\mnras},
         year = 2020,
        month = aug,
       volume = {496},
       number = {4},
        pages = {4468-4487},
          doi = {10.1093/mnras/staa1646},
archivePrefix = {arXiv},
       eprint = {2002.03867},
 primaryClass = {astro-ph.CO},
       adsurl = {https://ui.adsabs.harvard.edu/abs/2020MNRAS.496.4468S}
}

\appendix
\section{Effect of the angular median on stellar profiles}\label{sec:appendix-a}
\begin{figure}
    \centering
    \includegraphics[width=0.8\linewidth]{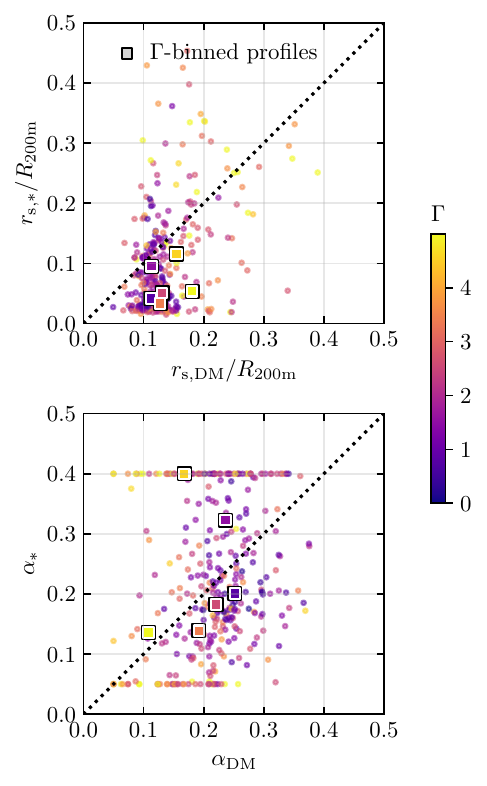}
    \caption{Comparison between $r_{\rm s}$ and $\alpha$ obtained from fitting the dark matter and stellar orbiting profiles. The points are colored by accretion rate (Equation \ref{eq:accretion-rate}). The square points are the result of fitting the $\Gamma$-binned median profiles.}
    \label{fig:rs-alpha-stars-vs-dm}
\end{figure}
\begin{figure*}
    \centering
    \includegraphics[width=0.8\linewidth]{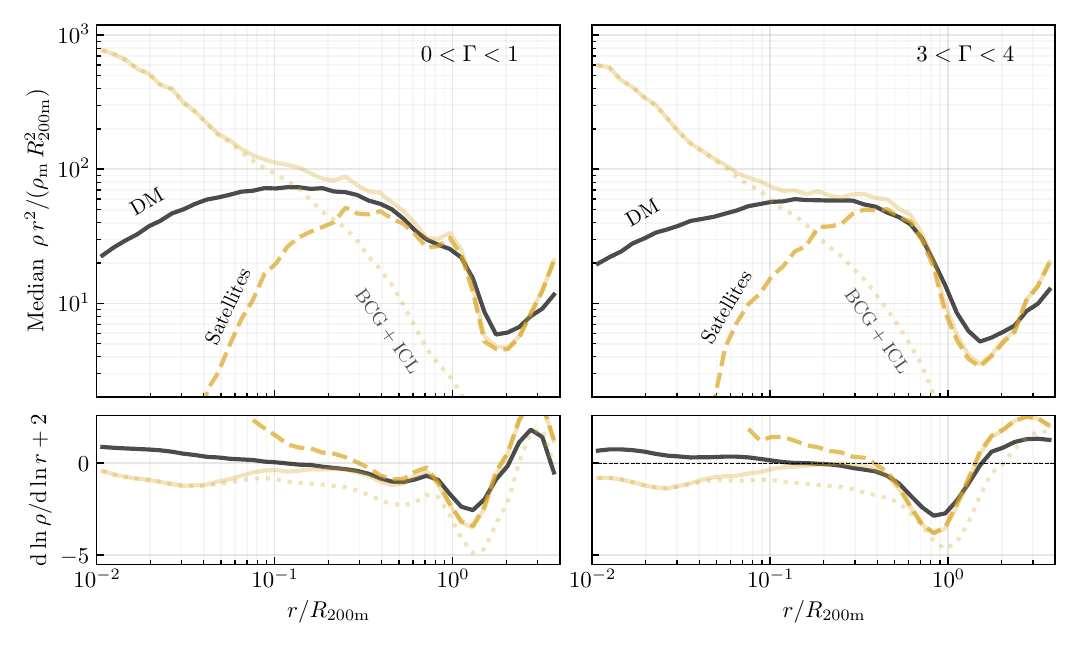}
    \caption{The same as Figure \ref{fig:DM-stellar-profile-comparison-angmean} but using angular-median density profiles.}
    \label{fig:DM-stellar-profile-comparison-angmed}
\end{figure*}
As described in \S\ref{sec:density_profiles}, the angular median method we use to calculate density profiles is much more effective at suppressing local fluctuations in density due to anisotropic substructure than the spherical average. This is because the median is much less sensitive to large outliers than the mean. Naturally, this affects the shape of the density profile and hence the parameters obtained from fitting.

Figure \ref{fig:rs-alpha-stars-vs-dm} compares the values of $r_{\rm s}$ and $\alpha$ determined from fitting the stellar and dark matter orbiting profiles. Unlike the parameters of the truncation term, $r_{\rm s}$ and $\alpha$ do not agree well between the stars and dark matter, despite what might be expected given the results presented in \S\ref{sec:inner-profile}. However, this disagreement may be partly understood by looking at the equivalent of Figure \ref{fig:DM-stellar-profile-comparison-angmean} for the angular median profiles, shown in Figure \ref{fig:DM-stellar-profile-comparison-angmed}. It is immediately clear that the density profile of stars in satellites is heavily suppressed---particularly around the dark matter $r_{\rm s}$. This is the expected effect of the angular median. However, the consequence is that the peak of the satellite profile is shifted and concealed by the BCG at all accretion rates, making the $r_{\rm s}$ measurement unreliable and disconnected from that of the dark matter. Visual inspection shows that this is reflected in the individual profiles, though there are exceptions.

\section{Satellite number density profiles}\label{sec:appendix-b}
\begin{figure}
    \centering
    \includegraphics[width=1\linewidth]{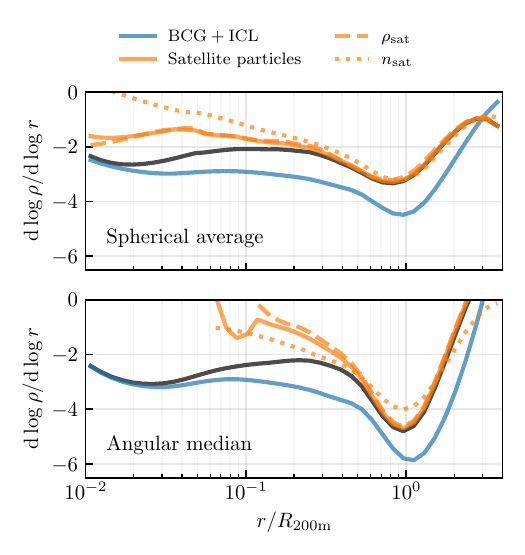}
    \caption{The mean stellar density slope profile of all \texttt{GIZMO\_7k} clusters for several different definitions, where the density profiles are calculated using a spherical average (top) and the angular median described in \S\ref{sec:density_profiles} (bottom). The black curve is the total density of stars, the solid orange and blue curves are the density of particles contained inside satellites and outside satellites (i.e. BCG+ICL), respectively. The dashed orange curve is the density slope of satellites, calculated by treating the satellites as point particles with mass equal to their stellar mass (as defined by \textsc{ahf}). Finally, the orange dotted curve is the slope of the number density profile of satellites.}
    \label{fig:satellite-icl-slopes}
\end{figure}
Given the challenges involved in observing the stellar density profile, it is worthwhile to assess how well we expect our results apply to galaxy number density profiles, which are more readily observable.

The top panel of Figure \ref{fig:satellite-icl-slopes} compares the slope of the mean total mass density of all the \texttt{GIZMO\_7k} clusters (black) to that of the number density of satellites $n_{\rm sat}$ (dotted orange), where the individual profiles have been calculated using a spherical average. We also show the slope of the density profile of particles in satellites (solid orange), of the smooth BCG+ICL (i.e. all other particles; blue), and of the density $\rho_{\rm sat}$ calculated by treating the satellites as point particles with mass equal to their stellar mass (as defined by \textsc{ahf}; dashed orange). All three profiles that are calculated using the satellites, regardless of the method, agree well with the total profile beyond $\sim 0.3R_{200{\rm m}}$. This is not unexpected given that the stellar mass outside the central region of clusters is dominated by stars bound in satellites. Meanwhile, the ICL profile is significantly different from the others, being steeper and having a slightly larger $r_{\rm steep}$. These results suggest that spherically-averaged satellite number density profiles accurately trace the total density profile around the splashback region.

The bottom panel of Figure \ref{fig:satellite-icl-slopes} is the same as the top panel, except that the profiles being averaged are calculated using the angular median method described in \S\ref{sec:density_profiles}. The angular median shifts $r_{\rm steep}$ slightly outwards but preserves the agreement between the profiles. It also makes the steepening feature of the profiles sharper, though this effect is not as significant for the number density. This is because the number density profiles effectively assign all satellites the same unit mass, which reduces the anistropy of the mass distribution in each radial shell. The angular median therefore has a significantly smaller effect on the number density than on the more anisotropic total mass density. However, this can (in principle) be mitigated by weighting the satellites by their stellar mass, as shown by the agreement between $\rho_{\rm sat}$ and total density slope around the splashback region. The sharpening of the splashback feature by the angular median may make it easier to identify in observations---even for clusters that don't show a visible feature in the satellite number density.

\bsp	
\label{lastpage}
\end{document}